\documentclass[12pt]{elsarticle}
\usepackage[utf8]{inputenc}
\usepackage{graphicx}
\usepackage{subfigure}

\usepackage{amssymb}
\usepackage{amsthm}
\usepackage{url}
\usepackage{amsmath}
\usepackage{subfigure}
\usepackage{textcomp}
\usepackage{xcolor}

\usepackage{fancyhdr}
\graphicspath{{figs/}}

\journal{BioSystems}

\begin{document}

\begin{frontmatter}

\title{Towards proteinoid computers. Hypothesis paper.}
\author{Andrew Adamatzky}

\address{Unconventional Computing Laboratory, UWE, Bristol, UK \\andrew.adamatzky@uwe.ac.uk}

\begin{abstract}
\noindent
Proteinoids --- thermal proteins ---are produced by heating amino acids to their melting point and initiation of polymerisation to produce polymeric chains. Proteinoids swell in aqueous solution into hollow microspheres. The  proteinoid microspheres produce endogenous burst of electrical potential spikes and change patterns of their electrical activity in response to illumination. The microspheres can interconnect by pores and tubes and form networks with a programmable growth. We speculate on how ensembles of the proteinoid microspheres can be developed into unconventional computing devices. 
\end{abstract}

\begin{keyword}
  thermal proteins, proteinoids, microspheres, unconventional computing
\end{keyword}

\end{frontmatter}


\section{Introduction}

The resources and methodologies used to manufacture electronic devices raise urgent questions about the negative environmental impacts of the manufacture, use, and disposal of electronic devices~\cite{cenci2021eco}. The use of organic materials to build electronic devices may offer an eco-friendly and affordable approach to growing our electronic world~\cite{han2020advanced,li2020biodegradable,wu2021biodegradable}. New organic computing substrates might create novel properties impossible to replicate with silicon, expanding the world of computing and electronics in ways unimaginable until now~\cite{lee2017toward,chang2017circuits,van2018organic,friederich2019toward}. The sister field of organic electronics is the unconventional computing which uncovers novel principles of efficient information processing in physical, chemical, biological systems, to develop novel computing substrates, algorithms, architectures~\cite{adamatzky2016advances,adamatzky2021thoughts}. Whilst there are many prototypes of organic electronic devices~\cite{ji2019recent,fahlman2019interfaces,mao2019bio,matsui2019flexible} few if any of them show a substantial degrees of stability or biocompatibility~\cite{feron2018organic}. This is why we propose to explore a unique class of organic devices --- thermal proteins~\cite{fox1992thermal} --- as a substrate and an architecture for future non-silicon massive parallel computers. We will discuss on how to hybridise the unconventional computing and organic electronics to design the prototype of a unique micro-scale proto-neuromorphic network of hollow protein microspheres.

\section{Proteinoids} 

\begin{figure}[!tbp]
\centering
\subfigure[]{\includegraphics[width=0.49\textwidth]{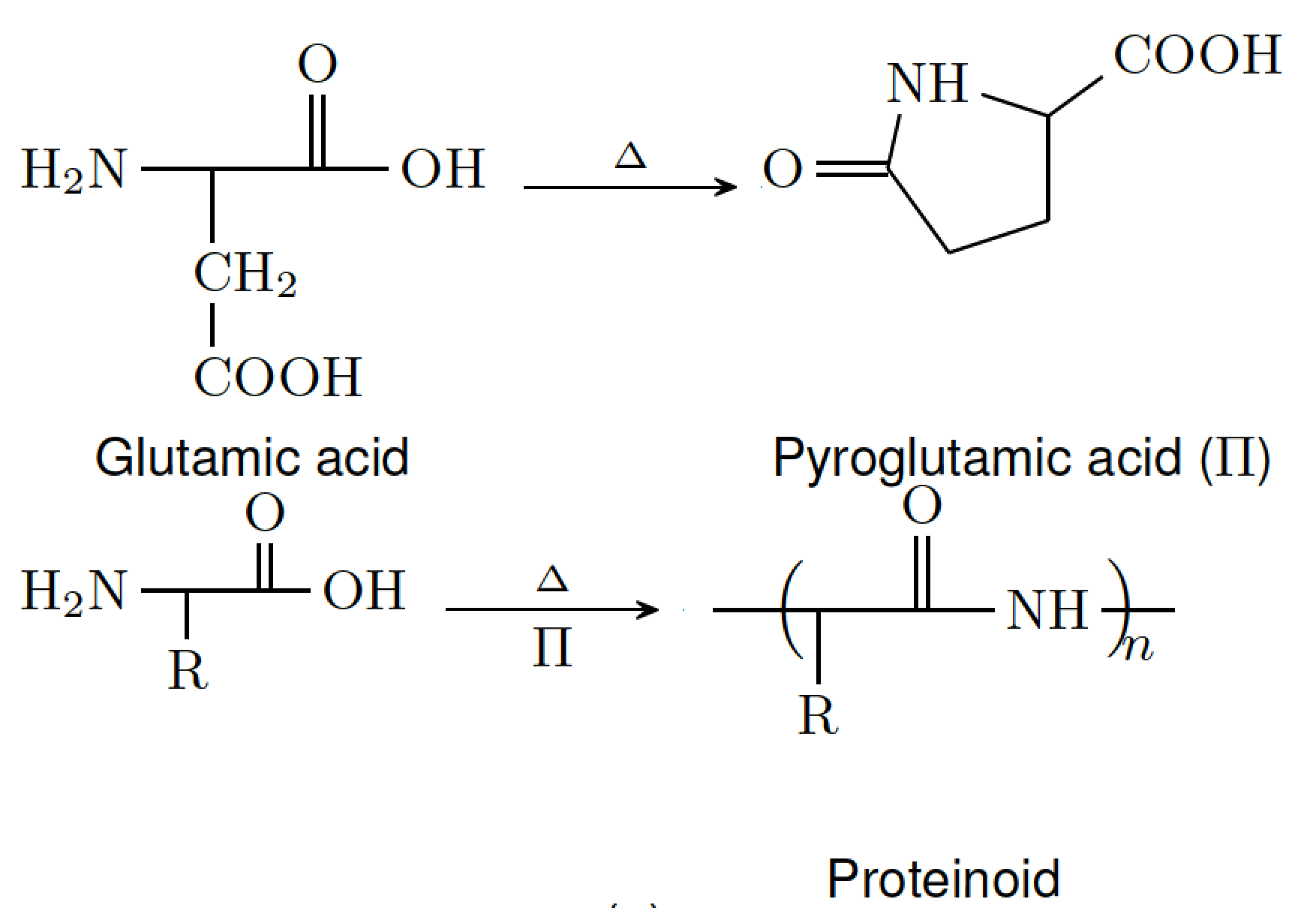}}
\subfigure[]{\includegraphics[width=0.49\textwidth]{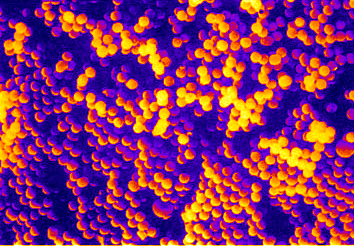}}
\subfigure[]{\includegraphics[width=0.05\textwidth]{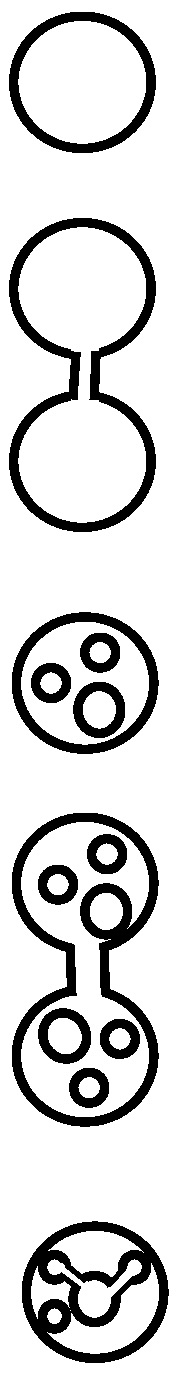}}
\subfigure[]{\includegraphics[width=0.49\textwidth]{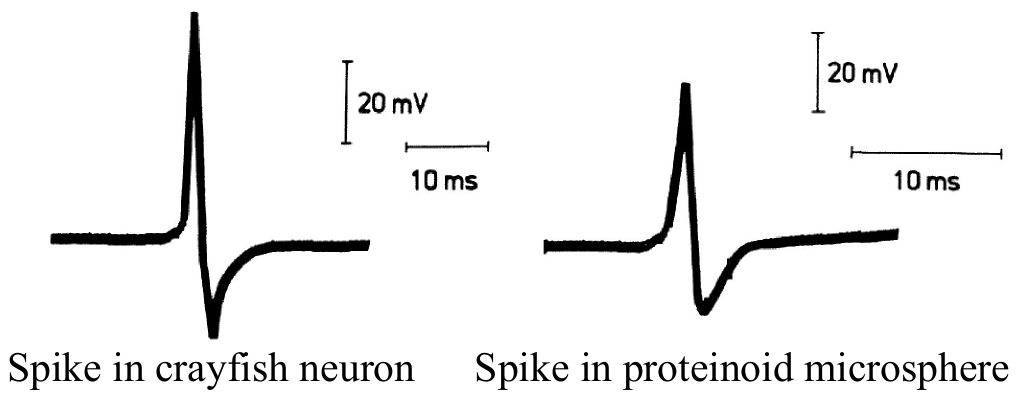}}
\caption{
(a)~A scheme of thermal polymerisation of amino acids through amino acid catalysis. Modified from~\cite{kolitz2014engineering}. 
(b) SEM photo of the proteinoid microspheres. Modified from~\cite{fox1995experimental}.  
(c)~Families of microspheres.
(d)~Action potentials from crayfish stretch receptor neuron and from proteinoid microsphere. Modified from~\cite{fox1995experimental}.
}
\label{scheme1}
\end{figure}

Thermal proteins (proteinoids)~\cite{fox1992thermal} are produced by heating amino acids to their melting point and initiation of polymerisation to produce polymeric chains (Fig.~\ref{scheme1}a). The polymerisation happens at 160–200~\textsuperscript{o}C, in absence of a solvent, an initiator or a catalyst,  in an inert atmosphere. The tri-functional amino acids, e.g. glutamic or aspartic acid or lysine, undergo cyclisation at high temperature and become solvents and initiators of polymerisation for other amino acids~\cite{harada1958thermal,fox1992thermal}.  It is possible to produce proteinoids that are either acidic or basic via this simple thermal condensation reaction. A proteinoid can be swelled in aqueous solution at moderate temperatures (c. 50~C\textsuperscript{o}) forming a structure known as a microsphere~\cite{fox1992thermal} (Fig.~\ref{scheme1}b). The microspheres are hollow, usually filled with aqueous solution.

 The growth of micro spheres is programmable. The sizes of the microspheres, 20$~\mu$m to 200~$\mu$m, can be programmed by selecting subsets of amino acids an thermal regimes~\cite{fox1992thermal}. 
Proteinoid microspheres of dominant hydrophobic constitution form `gap junctions' (Fig.~\ref{scheme1}c), sprout axon-like outgrowths, and form dendritic networks spontaneously. The microspheres can be embedded into larger microspheres. A communication between microspheres occurs via internal material exchange through the junctions~\cite{hsu1971conjugation}. The proteinoids and dried microspheres remain stable for up 10 years~\cite{Rohlfing1970998,przybylski1982membrane}.

\section{Electrical activity: membrane potential and oscillations}

The proteinoid microspheres maintain a steady state membrane potential 20~mV to 70~mV without any stimulating current and some microspheres in the population display the opposite polarization steadily~\cite{przybylski1985excitable}. Electrical membrane potentials, oscillations, and action potentials (Fig.~\ref{scheme1}d) are observed in the microspheres impaled with microelectrodes. These microspheres exhibit action-potential like spikes. The electrical activity of the microspheres also includes spontaneous bursts of electrical potential (flip-flops), and miniature potential activities at flopped phases~\cite{ishima1981electrical}. Although effects are of greater magnitude when the microspheres contain glycerol  lecithin, purely synthetic microspheres can attain 20~mV membrane potential~\cite{przybylski1982membrane}. In pure 20~$\mu$m microspheres  an amplitude is 20~mV, in 200~$\mu$m microspheres with lecithin 70~mV. The amplitude of spiking is regular in the phospholipid-free microspheres~\cite{przybylski1982membrane}. Membrane, action, and oscillatory potentials recorded from the microspheres composed of thermal protein, glycerol, and lecithin~\cite{ishima1981electrical,przybylski1982membrane} are observed for several days~\cite{bi1994evidence}. The microspheres remain stable~\cite{Rohlfing1970998} in aqua at pHs above 7.0\textsuperscript{o}C and continue oscillating for weeks~\cite{przybylski1985excitable}. 

Exact mechanisms of the proteinoid excitability are not fully agreed on. The physical conditions of the excitability are outlined in~\cite{przybylski1984physical}. One of the theoretical models is based on the strength of the coupling between basic and acidic proteinoids where the change abruptly with time due to the material flow~\cite{matsuno1984electrical}. Other models worse considering would be the classical model of ionic gradients established via channels~\cite{kimizuka1964ion} and a generation of the membrane potential by the immobilisation of mobile ions, where the ions absorb on the surface of the proteinoid microspheres, thus generating the nonzero membrane potential~\cite{tamagawa2015membrane}.  Also we could not rule out a salt water oscillator model~\cite{nakata1998self} which might be valid even in case where pores in the microspheres are passive.

\section{Response to stimulation}

\begin{figure}[!tbp]
    \centering
    \subfigure[]{\includegraphics[width=0.6\textwidth]{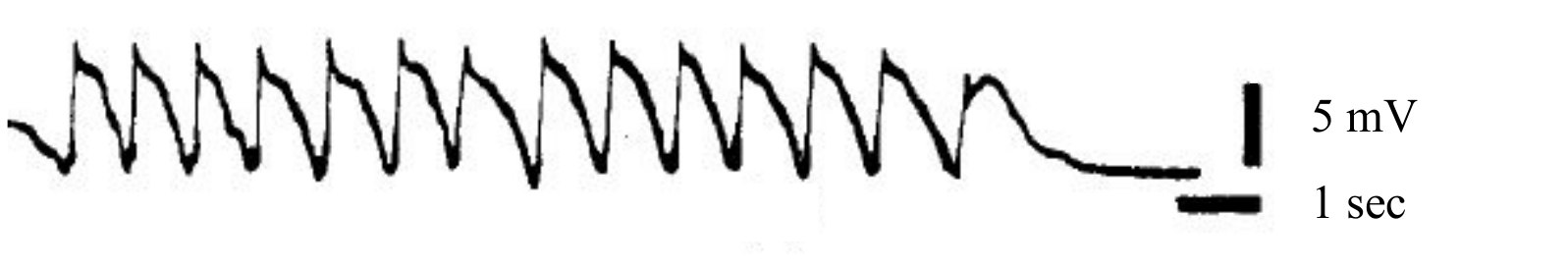}}
\subfigure[]{\includegraphics[width=0.7\textwidth]{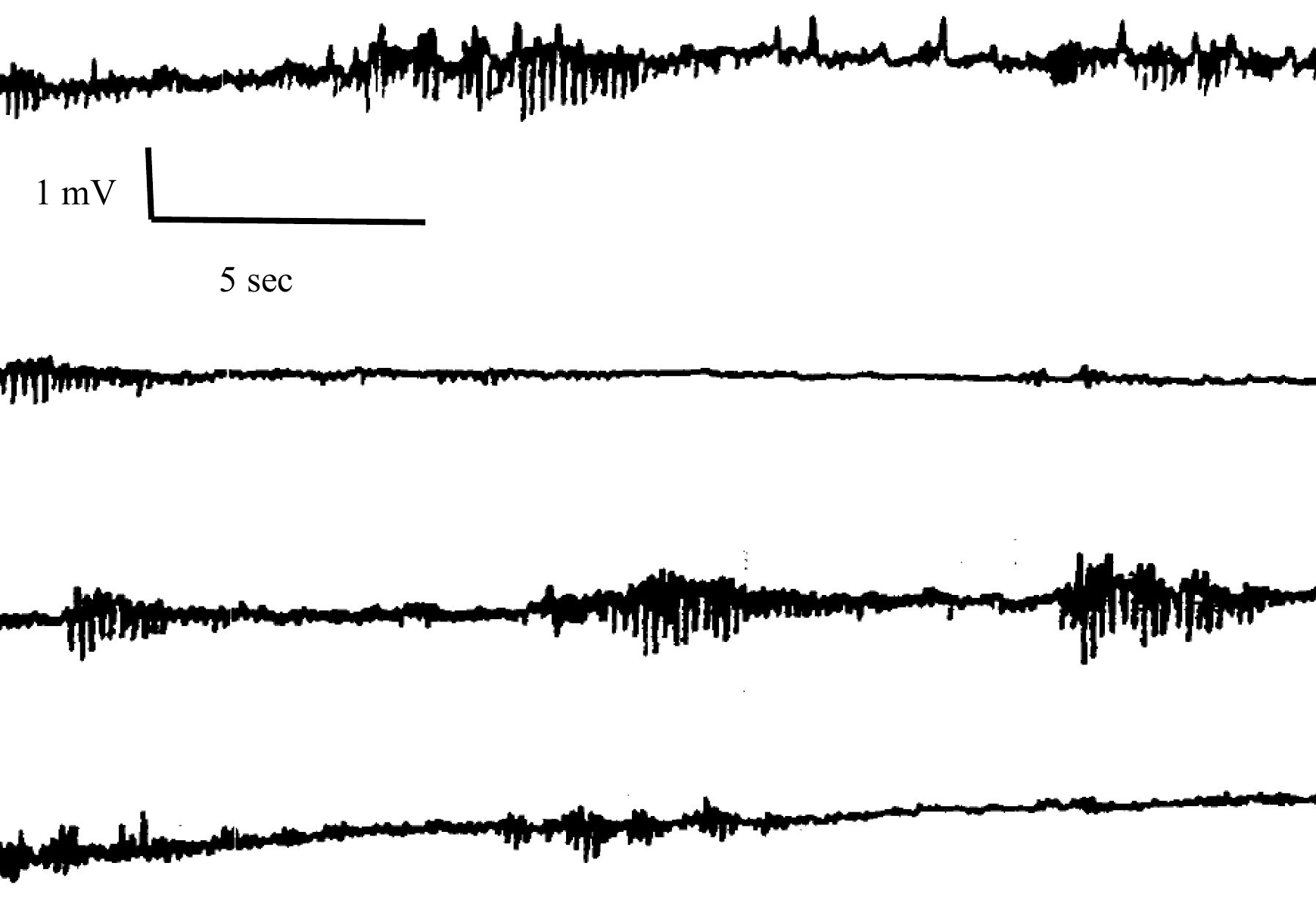}}
    \caption{Electrical activity of proteinoid microspheres. 
    (a)~Endogenous activity. Modified from~\cite{ishima1981electrical}. 
    (b)~Electrical discharges of the proteinoid sphere made of poly(asp:glu:arg) in response to illumination. Modified from~\cite{przybylski1985excitable}. }
    \label{fig:spikeTrains}
\end{figure}

The proteinoids are photosensitive. When the aqueous solution of the polymer is illuminated, a photovoltaic effect is recorded.  By varying composition of the amino acids it is possible to get either electron-donor or electron-acceptor~\cite{przybylski1983towards}. The spontaneous electrical discharges of the microspheres have been recorded during moderate illumination (Fig.~\ref{scheme1}e and \ref{fig:spikeTrains}). When the light was off, the gradual slowing of discharges up to complete cessation was observed. While illuminated by  intense light, the  oscillations stop on the higher level of the  potential~\cite{ishima1981electrical}.

\section{Pathways to prototyping}

Proteinoid microspheres are electrically active. They produce endogenous burst of electrical potential spikes. They can interconnect by pores and tubes and form networks. They respond to stimulation with trains of spikes. Their growth is programmable. The electrical properties, coupled with the connectional properties of the proteinoid microspheres invite us to engage in engineering an artificial brain through the artificial neuron and the self-organising properties that were employed by evolutionary processes. Thus the ensembles of the microspheres can be considered as networks of coupled oscillators or a  physical protoneural network (if junctions between microspheres are dynamically modifiable). We propose to investigate how these protoneural networks process information and to evaluate a potential of the networks to be the computing devices.

To prove that computing devices can be implemented using proteinoid microspheres one needs to develop theoretical designs and experimental laboratory prototypes of the sensing and information processing devices from ensembles of the microsphere. This can be done in three steps. First, we will develop an experimental protocol for producing ensembles of proteinoids and recording their electrical activity. Second, we will  classify patterns of endogenous electrical activity of the ensembles of proteinoids and  catalogue stimuli-to-responses electrical patterns. Third, we will implement mappings of binary strings by ensembles of proteinoids and characterise families of logical circuits implementable by the ensembles. 

The proteinoids can be prepared by heating mixtures of amino acids at 190\textsuperscript{o}C under a nitrogen blanket, mechanically stirred with water and filtered~\cite{przybylski1985excitable,kumar1998preparation,sasson2020engineering,kolitz2015engineering}.  Microspheres assembled from copoly(asp, glu), copoly(asp, glu) complexed with copoly(asp, glu, arg, his). It would be also advantageous to study leucine-rich, threonine-rich, prolinerich, and leucine-rich proteinoid: these water-soluble proteinoids formed semipermeable membranes with lecithin~\cite{ishima1981electrical}. Among these, the leucine-rich  proteinoid alone produce pronounced electrical activities~\cite{ishima1981electrical}. 

Intra-microsphere recording can be performed using capillary electrodes and recordings from ensembles of microspheres will be done via arrays of iridium coated electrodes and using voltage-sensitive dies. Recording of electrical behaviour to be done with Pico ADC-24 loggers fed from micro-electrode array, stimulation with impulse/function generators and cold light sources. The electronic characteristics will be assessed with semiconductor characterisation systems and sourcemeters. Ensembles recorded would consist of up to 40K microspheres in a monolayer or up to 200-300K microspheres in a multilayer, with linear size of the ensembles up to 4~mm. The electrical spiking activity of the proteinoid ensembles might be highly variable compared to neural activity and therefore might not be analysed by standard tools from the neuroscience. We propose to adopt the original techniques for detecting and classifying the spiking activity of fungi~\cite{dehshibi2021electrical}.

In order to gain insight into the underlying mechanism of the electrical behaviour of the proteinoid spheres, the following experimental data need to be collected: light absorbance, light effect on electrical activity of the spheres, membrane polarisation, current-voltage characteristics, and negative resistance of the polymer, as well as of the membrane, the relationship between copolyamino acid constitution and electrical readout patterns. 

The first step towards design of sensing and computing devices from the proteinoid microspheres would be in analysing the endogenous electrical activity of the microspheres ensembles. Such an analysis can be done via studying origins of spontaneous discharges and classification of  (ir)regularly oscillating potential dynamics. It might be advantageous to  distinguish patterns of spontaneous activity formed by a single spike and different types of bursts according to intra-burst firing frequency and to classify the bursting activity by coefficients of variation of the periods between spikes and serial correlations coefficients of the inter-spike intervals~\cite{perkel1967neuronal,cocatre1992identification,aldridge1991temporal,dorval2008probability,lobov2020competitive}. We could evaluate  a likelihood of a synchronisation of bursts recorded from microspheres of the same cluster~\cite{zheng2008spatiotemporal,sun2011burst,ahn2020synchrony}, and an interplay with plasticity and internal bursting~\cite{slomowitz2015interplay,desai1999plasticity,aberg2020interplay}.

Developing a dynamical model of the growing and re-configurable 3D ensemble of proteinoids. The key component of the model is a process where the external stimuli received by a local domain of the ensemble is accumulated  and then transmitted, in the form of propagating excitation wave fronts (recorded as bursts of spikes) or phase shifts in oscillation frequency model parameters, e.g.  kinetic constants and the transport coefficients, will be obtained by the fitting procedure in laboratory experiments.  

The analysis of the endogenous activity would result in a dictionary of the patterns of endogenous spiking of proteinoid ensembles. This dictionary would give us a `ground mode' of electrical behaviour and would help to distinguish between electrical events happening in the undisturbed ensembles and those caused by external input stimuli.

A computation with the ensembles of proteinoid microspheres can be implemented in many ways. Based on our previous experience in realising computing schemes in various unconventional substrates, two most feasible approaches would be (1)~to adopt algorithms and prototypes of excitable medium computers and to (2)~employ techniques of reservoir computing. 

Ensembles of proteinoid microspheres can be seen as ensembles of microvolumes of an excitable medium, e.g. analogous to Belousov-Zhabotinsky (BZ) medium~\cite{belousov1959periodic, zhabotinsky1964periodic}. This analogy would allow us to nearly directly adapt all algorithms and prototyping designs of computing with BZ medium to the proteinoid computers. These include the image processes and memory devices~\cite{kuhnert1986new, kuhnert1989image, kaminaga2006reaction},  logical gates implemented in geometrically constrained BZ medium~\cite{steinbock1996chemical, sielewiesiuk2001logical}, approximation of shortest path by excitation waves~\cite{steinbock1995navigating, rambidi2001chemical, adamatzky2002collision}, memory in BZ micro-emulsion \cite{kaminaga2006reaction}, information coding with frequency of oscillations~\cite{gorecki2014information}, on-board controllers for robots~\cite{adamatzky2004experimental, yokoi2004excitable, DBLP:journals/ijuc/Vazquez-OteroFDD14}, chemical diodes~\cite{DBLP:journals/ijuc/IgarashiG11}, neuromorphic architectures~\cite{ gorecki2006information, gorecki2009information, stovold2012simulating, gentili2012belousov, takigawa2011dendritic, stovold2012simulating,  gruenert2015understanding} and associative memory \cite{stovold2016reaction,stovold2017associative}, wave-based counters~\cite{gorecki2003chemical}, logical gates~\cite{steinbock1996chemical, sielewiesiuk2001logical, adamatzky2004collision, adamatzky2007binary, toth2010simple, adamatzky2011towards,adamatzky2011polymorphic,stevens2012time}.

The excitable medium algorithms referenced above are strongly dependent on geometrical arrangement of input and output loci of the medium and, often, geometrical constraints on propagation of the excitation wave fronts. When a direct control of proteinoid microspheres' ensembles becomes unfeasible we could adopt a more relaxed approach of reconstructing mapping of binary strings by the ensembles with potential further extraction of the logical gates implemented.  This approach was pioneered in our numerical models of the computation on actin bundles and developmnet of an actin droplet machine~\cite{adamatzky2019actin}. 

A proteinoid machine is an ensemble of proteinoid microspheres interfaced with an arbitrary selected set of $k$ electrodes through which stimuli, binary strings of length $k$ represented by impulses generated on the electrodes, are applied and responses are recorded. The responses are recorded in a form of spikes and then converted to binary strings. The machine's state is a binary string of length $k$: if there is an impulse recorded on the $i$~th electrode, there is a `1' in the $i$~th position of the string, and `0' otherwise. 

All possible configurations of binary of length $k$ could be send to proteinoid ensembles. Thus, the ensemble will be implementing mapping $E: \{0,1\}^k \rightarrow \{0,1\}^k$.  We can consider $k=4, 8, 18, 32$ I/O setups. The mappings' graphs could be analysed using basic measure (in- and out-degrees distributions),  the length of the shortest paths, graph distance matrix, vertex eccentricity, graph radius, graph diameter,  node and edge connectivities, closeness centrality, degree  centrality, eigenvector centrality. The graphs can be compared using a set of metrics outlines in \cite{wills2020metrics}.

 From the mappings $E$ we can extract $k$-inputs-1-output functions. These functions will be analysed and classified by  weight, algebraic degree, nonlinearity, query complexity, circuit complexity, influence, sensitivity, block sensitivity and certificate complexity. Results will be in the behavioural characterisation of the finite state machine implementable by ensembles of proteinoid microspheres and families of logical functions implementable by the ensembles.
 
  The approach belongs to same family of computation outsourcing techniques as  \emph{in materio} computing~\cite{miller2002evolution,miller2014evolution,stepney2019co,miller2018materio,miller2019alchemy} and reservoir computing~\cite{verstraeten2007experimental,lukovsevivcius2009reservoir,dale2017reservoir,konkoli2018reservoir,dale2019substrate,tsakalos2021protein}.
 
 \section{Example. Computing Boolean functions on an ensemble of proteinoid microspheres.}
 
 In this example we will show how to realise Boolean gates with a disordered ensemble of proteinoid microspheres interfaced with an array of electrodes.
 
 \begin{figure}[!tbp]
     \centering
     \includegraphics[width=0.9\textwidth]{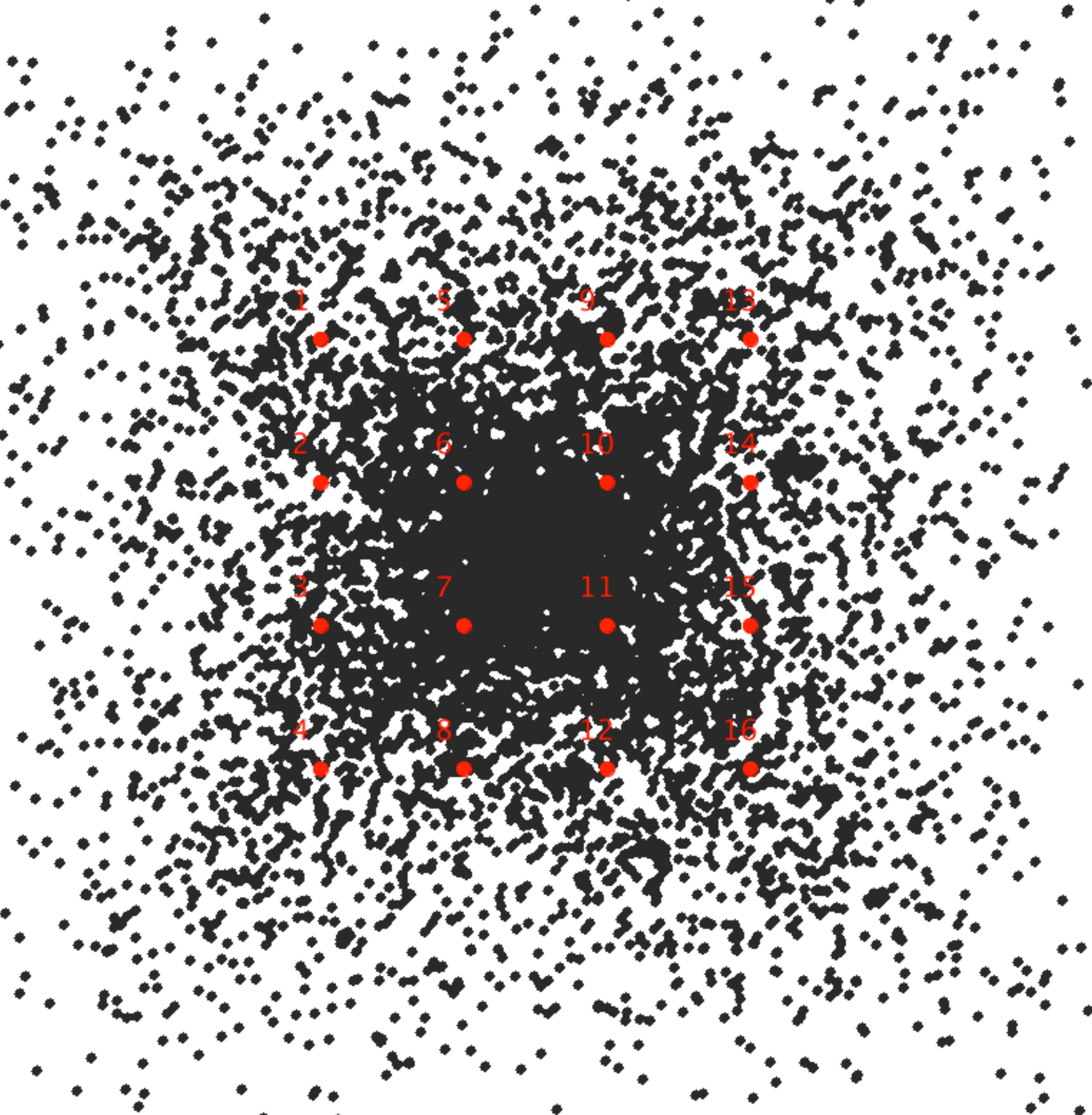}
     \caption{Configurations of the simulated ensemble of proteinoid microspheres, shown in black, and electrode array, shown in red. }
     \label{fig:ensemble}
 \end{figure}
 
 A simulated monolayer ensemble of proteinoid microspheres is generated by placing solid discs 7 pixels diameter on the finite lattice. To imitate higher density of the microsphere near the centre of the lattice we placed the discs with probabilities decreasing with increasing distance from the centre (Fig.~\ref{fig:ensemble}). 
 
 We used still image of the ensemble as a conductive template. The image of the ensemble was projected onto a $1000 \times 960$ nodes grid. 
The original image $M=(m_{ij})_{1 \leq j \leq n_i, 1 \leq j \leq n_j}$, $m_{ij} \in \{ r_{ij}, g_{ij}, b_{ij} \}$, where $n_i=1000$ and $n_j=960$, and $1 \leq r, g, b \leq 255$, was converted to a conductive matrix $C=(m_{ij})_{1 \leq i,j \leq n}$ derived from the image as follows: $m_{ij}=1$  if $r_{ij}<20$, $g_{ij}<20$ and $b_{ij}<20$. 

We simulated excitation of proteinoid microspheres with FitzHugh-Nagumo equations. The FitzHugh-Nagumo (FHN) equations~\cite{fitzhugh1961impulses,nagumo1962active,pertsov1993spiral} is a qualitative approximation of the Hodgkin-Huxley model~\cite{beeler1977reconstruction} of electrical activity of living cells:
\begin{eqnarray}
\frac{\partial v}{\partial t} & = & c_1 u (u-a) (1-u) - c_2 u v + I + D_u \nabla^2 \\
\frac{\partial v}{\partial t} & = & b (u - v),
\end{eqnarray}
where $u$ is a value of a trans-membrane potential, $v$ a variable accountable for a total slow ionic current, or a recovery variable responsible for a slow negative feedback, $I$ {is} a value of an external stimulation current. The current through intra-microsphere spaces is approximated by
$D_u \nabla^2$, where $D_u$ is a conductance. 

We integrated the system using the Euler method with the five-node Laplace operator, a time step $\Delta t=0.015$ and a grid point spacing $\Delta x = 2$, while other parameters were $D_u=1$, $a=0.13$, $b=0.013$, $c_1=0.26$. An excitability of the medium is controlled by parameter $c_2$. In the results presented $c_2=0.095$. Boundaries are considered to be impermeable: $\partial u/\partial \mathbf{n}=0$, where $\mathbf{n}$ is a vector normal to the boundary.

 \begin{figure}[!tbp]
     \centering
    \subfigure[$t=5000$]{\includegraphics[width=0.32\textwidth]{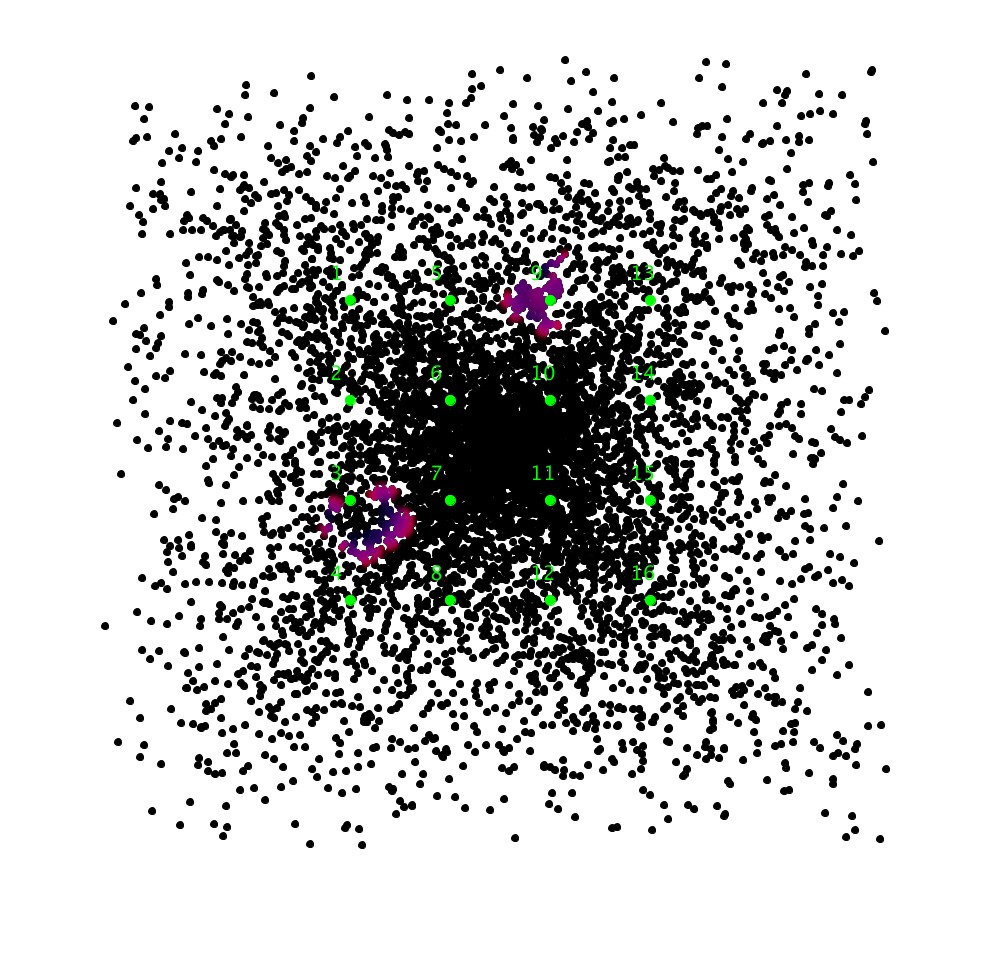}}
    \subfigure[$t=10000$]{\includegraphics[width=0.32\textwidth]{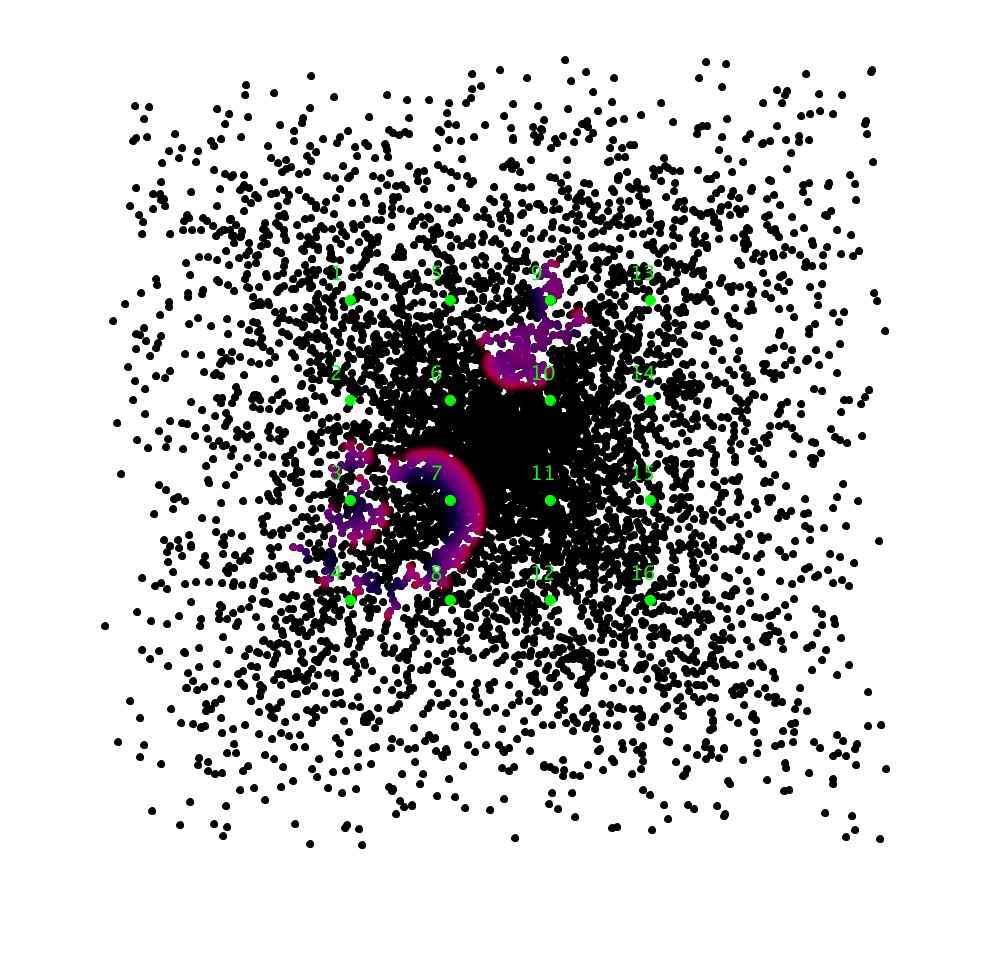}}
    \subfigure[$t=20000$]{\includegraphics[width=0.32\textwidth]{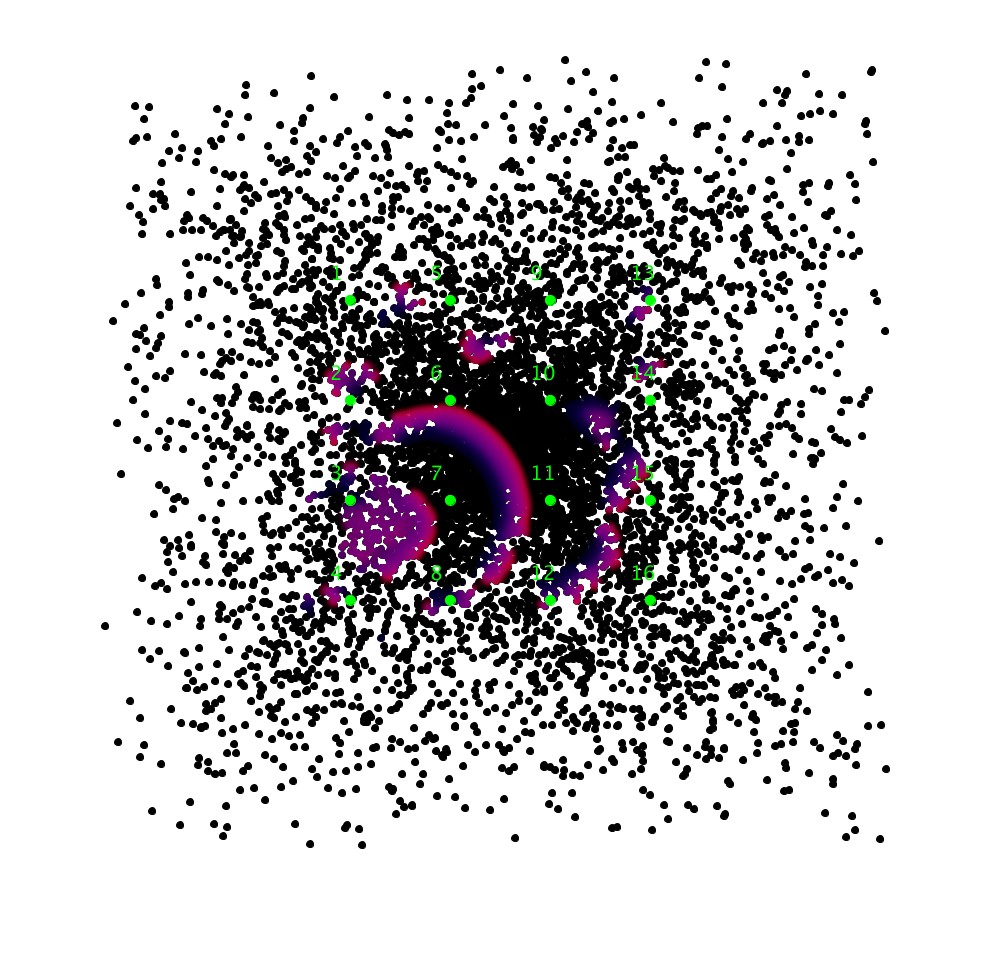}}
    \subfigure[$t=25000$]{\includegraphics[width=0.32\textwidth]{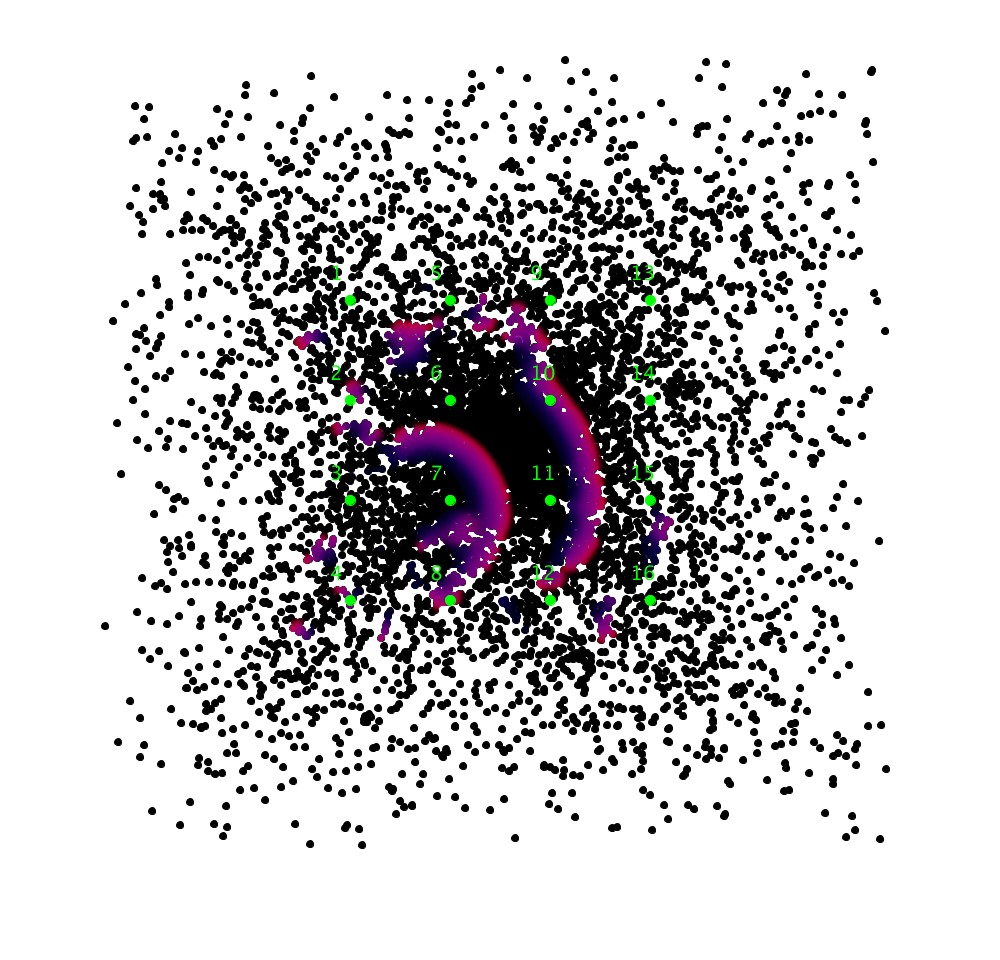}}
            \subfigure[$t=35000$]{\includegraphics[width=0.32\textwidth]{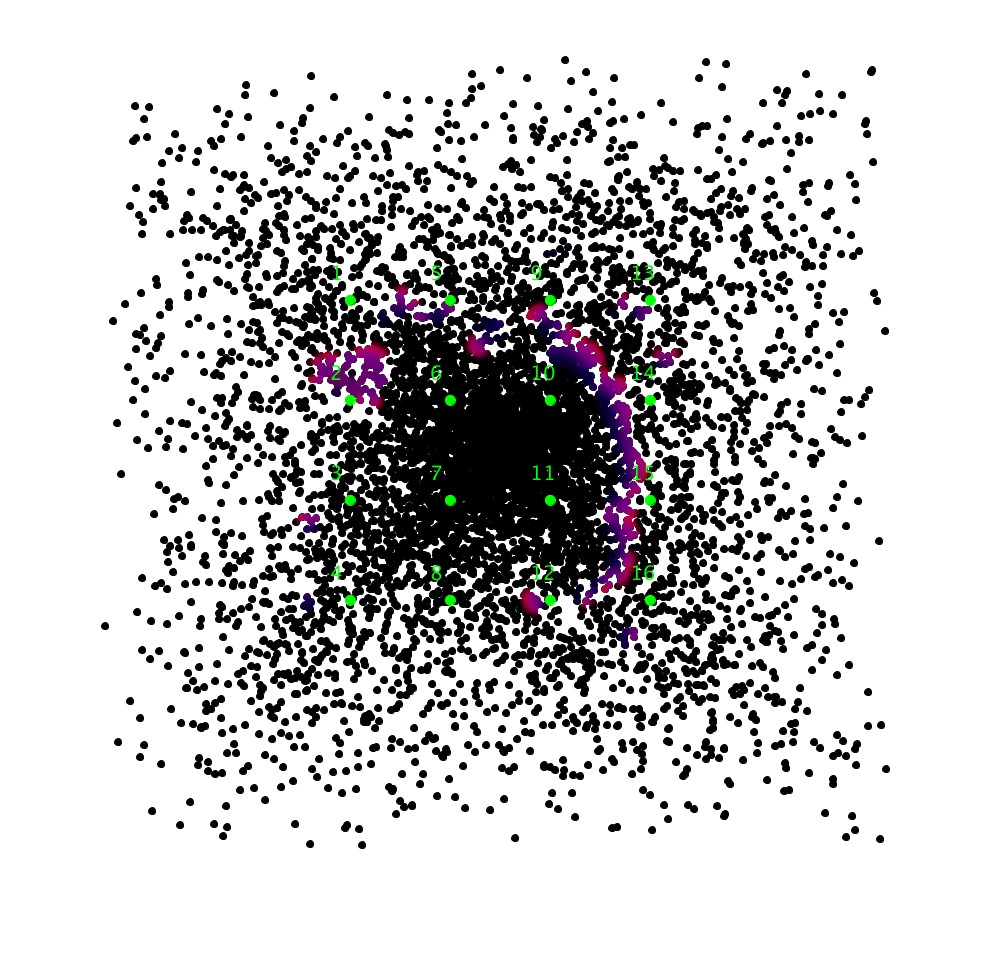}}
         \subfigure[$t=40000$]{\includegraphics[width=0.32\textwidth]{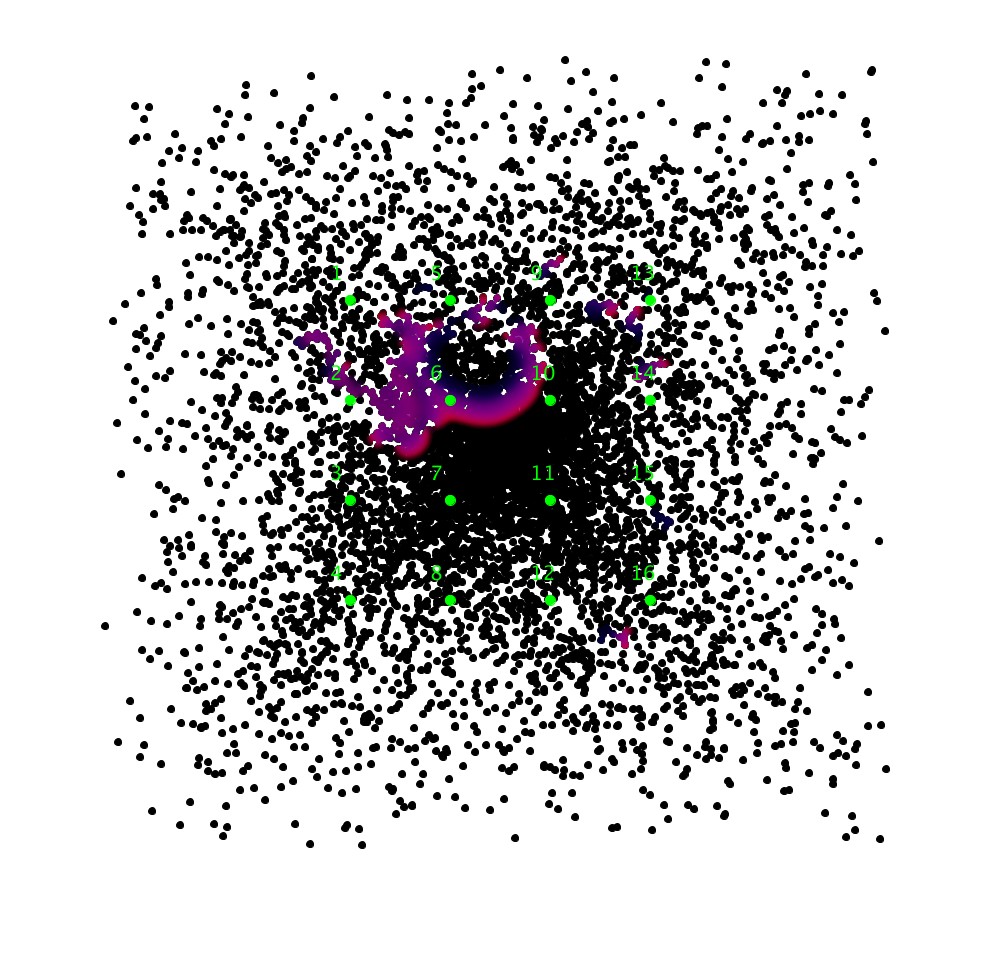}}
         \subfigure[$t=50000$]{\includegraphics[width=0.32\textwidth]{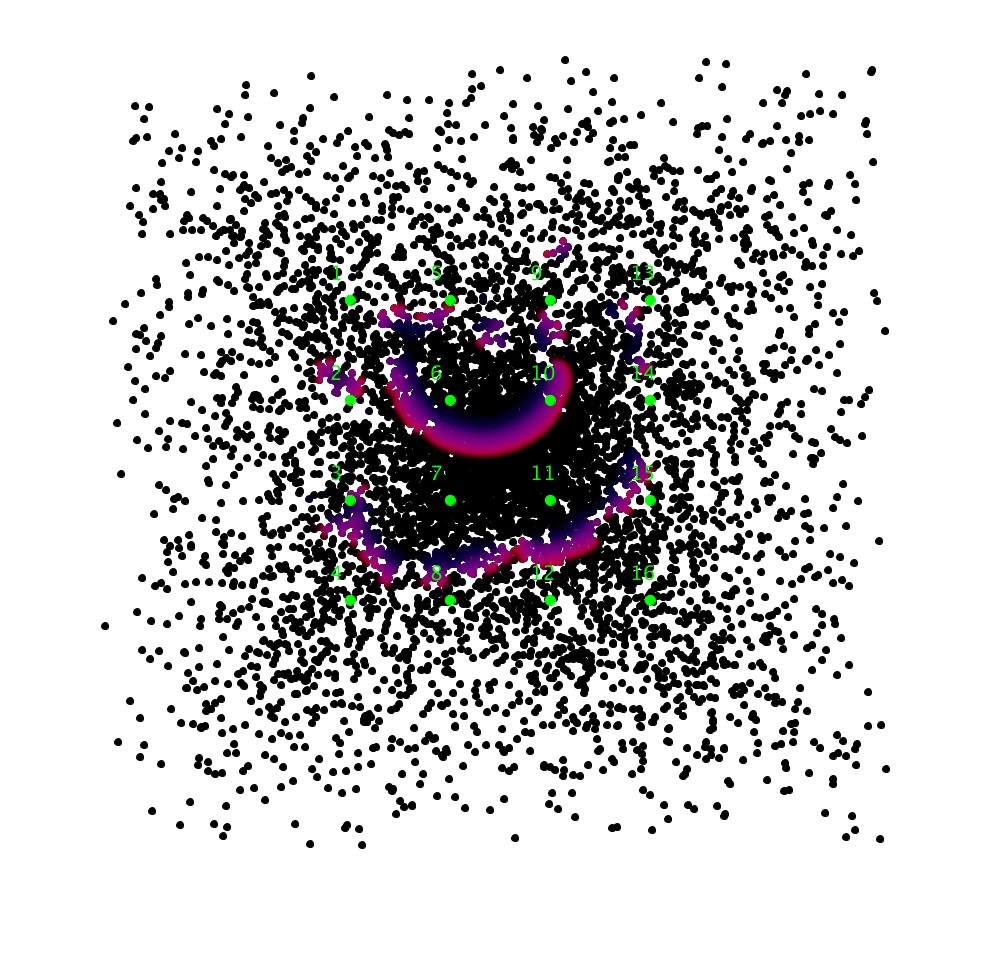}}
            \subfigure[$t=60000$]{\includegraphics[width=0.32\textwidth]{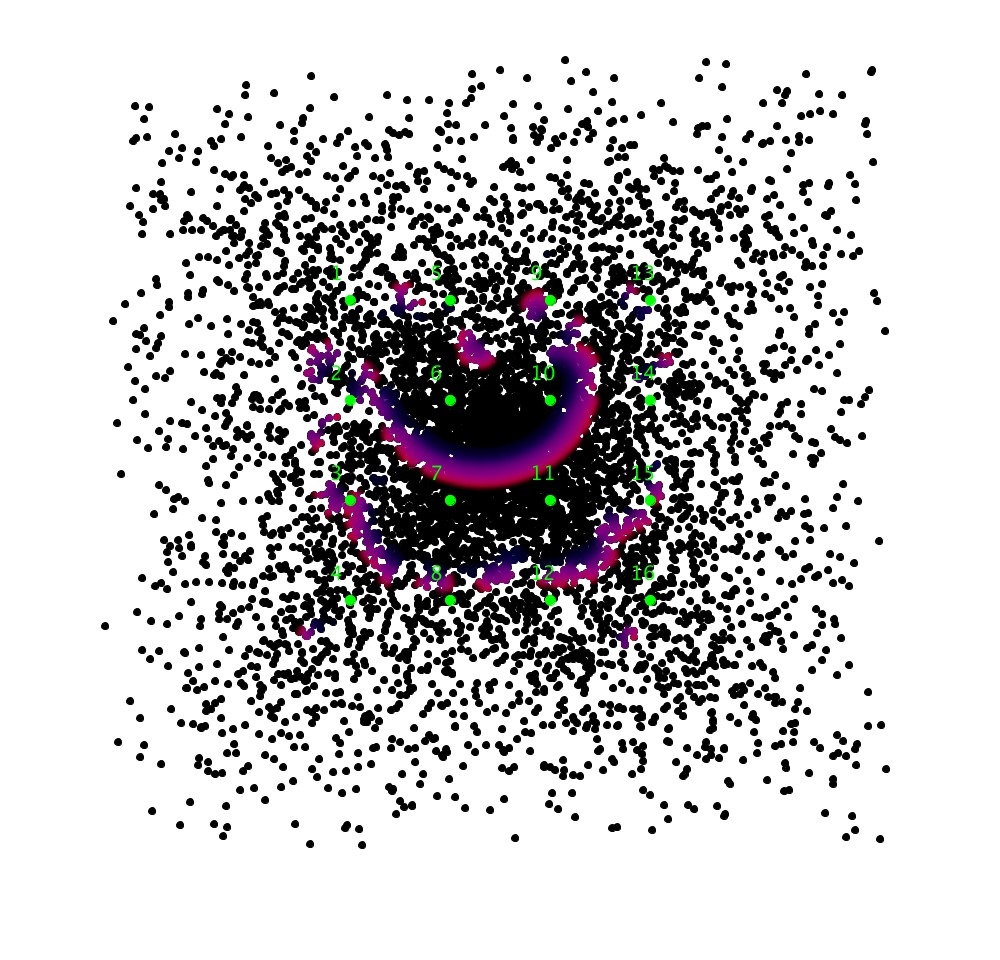}}
        \subfigure[$t=70000$]{\includegraphics[width=0.32\textwidth]{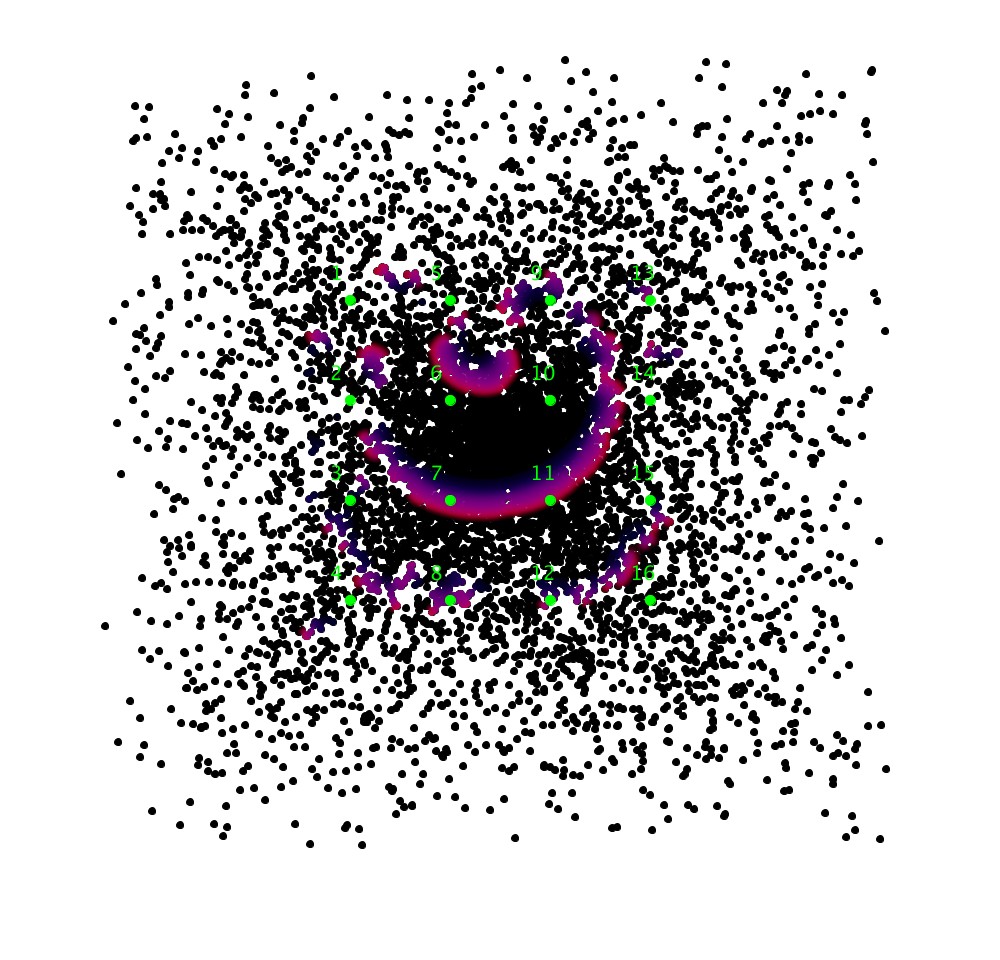}}
     \caption{Snapshots of excitation dynamics on a simulated ensemble of proteinoid microspheres. Excitation is stimulated by applying pulses to electrodes $E_9$ and $E_{3}$.}
     \label{fig:wavesSnapshots}
 \end{figure}

To show dynamics of excitation in the ensemble we simulated electrodes by calculating a potential $p^t_x$ at an electrode location $x$ as $p_x = \sum_{y: |x-y|<2} (u_x - v_x)$. Configuration of electrodes $1, \cdots, 16$ is shown in Fig.~\ref{fig:ensemble}. The configuration selected imitates a micro-electrode array.  Time-lapse snapshots provided in the paper were recorded at every 100\textsuperscript{th} time step, and we display sites with $u >0.04$; videos and figures were produced by saving a frame of the simulation every 100\textsuperscript{th} step of the numerical integration and assembling the saved frames into the video with a play rate of 30 fps. Videos are available at \url{https://doi.org/10.5281/zenodo.4884999}. Examples of the excitation dynamics are shown in Fig.~\ref{fig:wavesSnapshots}. 
 
 \begin{figure}[!tbp]
     \centering
\subfigure[]{\includegraphics[width=1\textwidth]{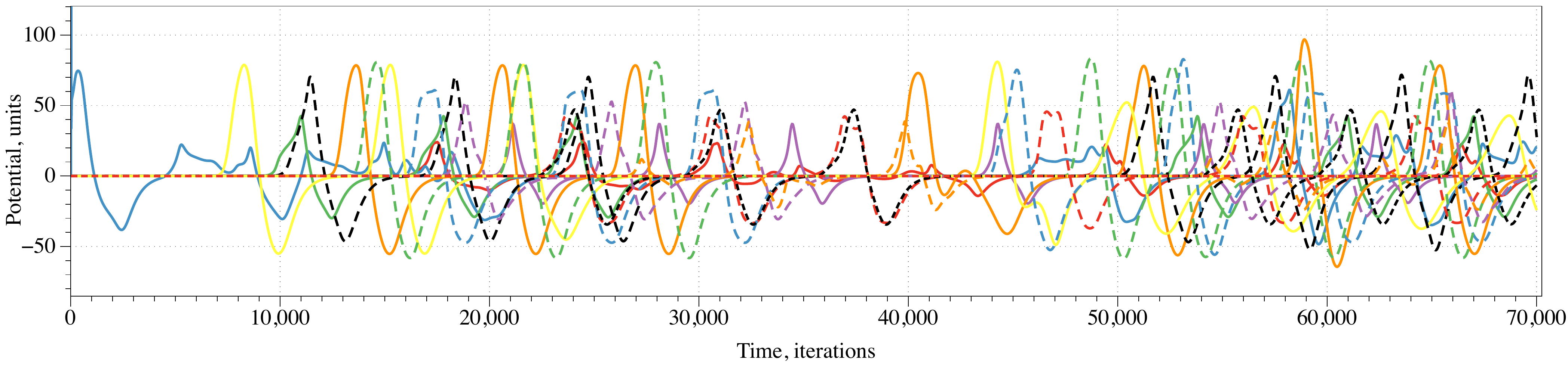}}
\subfigure[]{\includegraphics[width=1\textwidth]{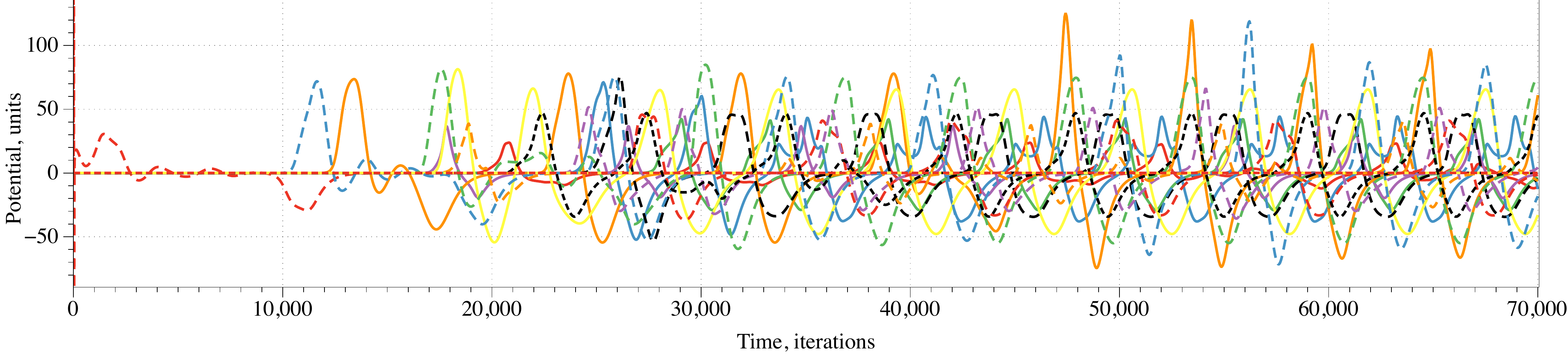}}
\subfigure[]{\includegraphics[width=1\textwidth]{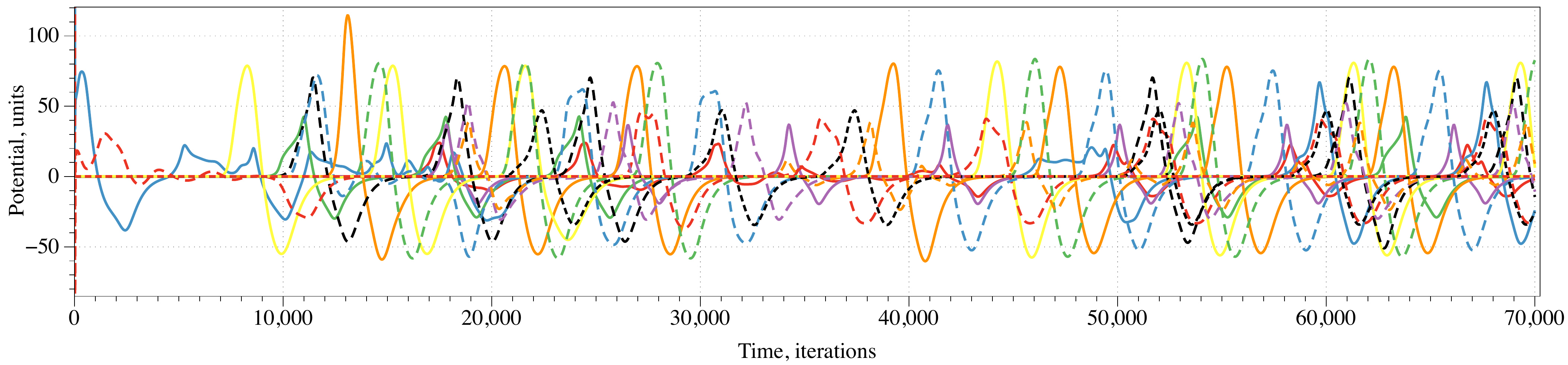}}
     \caption{Simulated electrical potential recorded at 16 electrodes for inputs (a)~(0,1), (b)~(1,0), (c)~(1,1). The inputs are represented by pulses generated at the electrodes $E_9$ and $E_{3}$: impulse on $E_9$ means input $(1,\star)$ and impulse on $E_3$ means input $(\star, 1)$.}
     \label{fig:potential}
 \end{figure}
 
 \begin{table}[!tbp]
    \centering
    \begin{tabular}{c|cc}
    spikes    & gate  & notations   \\  \hline
\includegraphics[scale=0.2]{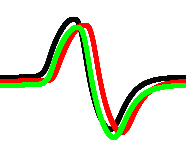}  & {\sc or} & $x+y$ \\
\includegraphics[scale=0.2]{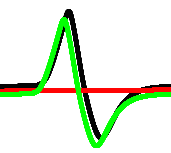}  & {\sc select} & $y$ \\
\includegraphics[scale=0.2]{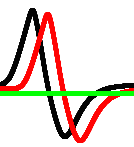}  & {\sc xor} & $x \oplus y$ \\
\includegraphics[scale=0.2]{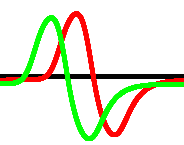}  & {\sc select} & $x$ \\
\includegraphics[scale=0.2]{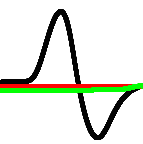}  & {\sc not-and} & $\overline{x}y$ \\
\includegraphics[scale=0.2]{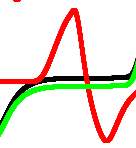}  & {\sc and-not} & $x\overline{y}$ \\
\includegraphics[scale=0.2]{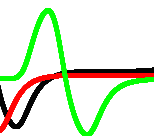}  & {\sc and } & $xy$ \\
    \end{tabular}
    \caption{Representation of gates by combinations of spikes. Black lines show the potential when the network was stimulated by input pair (01), red by (10) and green by (11). Adamatzky proposed this representation originally in \cite{adamatzky2019computing}.}
    \label{tab:spikes2gates}
\end{table}
 
 A spiking activity of the network shown in Fig.~\ref{fig:potential} in a response to stimulation, i.e. application of inputs $(0,1)$, $(1,0)$ and $(1,1)$ via impulses at the electrodes $E_9$ and $E_{3}$, recorded from electrodes $E_1, \cdots, E_{16}$. We assume that each spike represents logical {\sc True} and that spikes occurring within less than $2 \cdot 10^2$ iterations are simultaneous. Then a representation of gates by spikes and their combinations can be implemented as shown in Tab.~\ref{tab:spikes2gates}. By selecting specific intervals of recordings we can realise several gates in a single site of recording. In this particular case we assumed that spikes are separated if their occurrences are more than  $10^3$ iterations apart. 
 
 \begin{table}[!tbp]
     \centering
     \begin{tabular}{p{1cm}|ccccccc|p{1cm}}
$E$	&	$x+y$	&	$Sy$	&	$x\oplus y$	&	$Sx$	&	$\overline{x}y$	&	$x\overline{y}$	&	$xy$	&	\tiny{Total gates per electrode}	\\ \hline
1	&	0	&	0	&	0	&	0	&	0	&	0	&	0	&	0	\\
2	&	0	&	0	&	0	&	0	&	0	&	0	&	0	&	0	\\
3	&	0	&	0	&	0	&	0	&	0	&	0	&	0	&	0	\\
4	&	0	&	0	&	1	&	0	&	0	&	7	&	0	&	8	\\
5	&	0	&	4	&	1	&	0	&	0	&	6	&	1	&	12	\\
6	&	1	&	0	&	1	&	3	&	2	&	1	&	2	&	10	\\
7	&	0	&	2	&	1	&	5	&	1	&	2	&	0	&	11	\\
8	&	2	&	1	&	1	&	1	&	0	&	5	&	1	&	11	\\
9	&	1	&	3	&	0	&	1	&	2	&	5	&	0	&	12	\\
10	&	0	&	0	&	2	&	4	&	3	&	0	&	1	&	10	\\
11	&	2	&	1	&	0	&	4	&	4	&	2	&	0	&	13	\\
12	&	1	&	3	&	1	&	1	&	0	&	5	&	1	&	12	\\
13	&	0	&	2	&	0	&	3	&	0	&	5	&	0	&	10	\\
14	&	0	&	0	&	0	&	5	&	3	&	0	&	0	&	8	\\
15	&	0	&	0	&	0	&	0	&	0	&	0	&	0	&	0	\\
16	&	0	&	2	&	0	&	3	&	2	&	5	&	0	&	12	\\ \hline
\tiny{Total gates implemented}	&	7	&	18	&	8	&	30	&	17	&	43	&	6	&		
     \end{tabular}
     \caption{Number of gates discovered for each of 16 electrodes.}
     \label{tab:numberOfgates}
 \end{table}
 
We calculated frequencies of logical gates' implementable at the electrodes as follows.  For each of the recording sites, electrodes, we calculated a number of gates realised during $14.2 \cdot 10^4$ iterations (Tab.~\ref{tab:numberOfgates}). The simulated ensemble of proteinoid microspheres can implement two-inputs-two-outputs logical gates when we consider values of two recording electrodes at the same specified interval. For example, a one-bit half-adder: one output is {\sc and} and another output is {\sc xor}, and the Toffoli gate: one output is {\sc select} and another {\sc xor}.

\section{Discussion}

When thinking about proteinoid microspheres as proto-neurons~\cite{fox1995experimental,rizzotti1998did,matsuno2012molecular} one might ask a question `Why proteinoid microspheres not liposomes?'.
Vesicles with lipid membrane are widely studied in the context of artificial life~\cite{cans2003artificial,pohorille2002artificial,kamiya2017giant,zhang2008artificial,zhang2021fabrication}, drug delivery~\cite{litzinger1992phosphatodylethanolamine,antimisiaris2021overcoming,egorov2021robotics}. Moreover, as we demonstrated in~\cite{Nomura2020}, ensembles of liposomes made of phospholipids combined with viscous amphipathic molecules exhibit spike-like dynamics in electrical potential. However, there are crucial limitations of lipid bilayer membranes as follows. Phospholipid membranes are highly effective barriers to polar and charged molecules, necessitating complex protein channels and pumps to allow the exchange of these molecules with the external environment~\cite{meierhenrich2010origin,nielsen2009biomimetic}. Contemporary phospholipid membranes are non-permeable to essential molecules for life,
growth, and replication~\cite{sharov2016coenzyme,stano2017minimal}. There is no plausible way of lipid synthesis described under prebiotic conditions~\cite{schreiber2019prebiotic}. Thus, even though lipid vesicles are prominent models in synthetic protocell research, the conceptual links between  RNA replication and compartmentalisation strategies are not well developed~\cite{schreiber2019prebiotic}.

What would be potential breakthroughs and applications of proteinoid computing devices? Fundamental science breakthrough of proteinoid microspheres computing devices will be in the experimental laboratory prototyping of a protoneural network capable for distributed sensing and collective information processing. There are evidences that a first cell might have been similar to a proteinoid microsphere: artificial fossilization of the laboratory products closely resemble ancient microfossils from ancient strata~\cite{fox1995experimental}.  The findings on propagation of excitation waves in ensembles of proteinoid microspheres will open ways for studying coupling of biochemical and other phenomena in ways that cannot be accomplished with other models such as bilayer membranes. The proteinoids are sufficiently accurate prototypes of terrestrial protocells with bioelectrical properties~\cite{follmann1982deoxyribonucleotide,van2017origin}.
Having most characteristics of excitable cells proteinoid microspheres have been considered as protoneurons~\cite{przybylski1985excitable}. A consensus was achieved that proteinoids directly led into neurons, which then self-associated into brains~\cite{fox1992thermal}. The association of the microspheres is already evidenced~\cite{fox1992thermal}. 

Applied science domains for future developments in unconventional computing with proteinoids include biodegradable electronics, personalised healthcare, and biocompatible computing. By uncovering mechanisms of distributed sensing and information processing in ensemble of proteinoid microspheres we could establish a new paradigms and experimental protocols for non-silicon biodegradable liquid electronics, and monitoring and micro-intervention devices. These include information processing in natural systems, biophysics and soft matter, non-silicon electronics, synthetic biology, theory of computation. The ensembles of proteinoid microspheres could play a key role in novel developments in healthcare technologies and future therapies, especially delivery and release~\cite{quirk2010triggered}, due to their protein‐like nature, biocompatibility, non‐toxicity~\cite{kolitz2018recent}. Proteinoids are already used to encapsulate drugs, e.g. methotrexate~\cite{kumar1998preparation}, hydroxyapatite~\cite{tallawi2011proteinoid}, cholesterol~\cite{kokufuta1983factors}, drugs for osteogenic differentiation~\cite{jun2020role}. Proteinoid microspheres are proposed to be used for treatments of wounds by delivering a sequestered reagent in a controlled manner~\cite{quirk2010triggered}. Bringing here knowledge of enzymatic logic interfaced with biosensor and drug delivery system~\cite{zavalov2012enzyme,privman2018enzyme}, we envisage that ensembles of proteinoid microspheres will be used as embedded controllers in  implantable drug delivery systems. The networks of proteinoid microspheres will exemplify bio-compatible organic electronic devices. 

The proteinoids are neurotrophic~\cite{vaughan1987thermal}, they might even delay ageing of neurons~\cite{fox1992thermal}. Therefore there is a high potential for the proteinoid computers to be implantable into living brains to act as augmenting, monitoring or prosthetic devices.


\begin{thebibliography}{100}

\bibitem{aberg2020interplay}
Kristoffer~Carl Aberg, Emily~Elizabeth Kramer, and Sophie Schwartz.
\newblock Interplay between midbrain and dorsal anterior cingulate regions
  arbitrates lingering reward effects on memory encoding.
\newblock {\em Nature communications}, 11(1):1--14, 2020.

\bibitem{adamatzky2004collision}
Andrew Adamatzky.
\newblock Collision-based computing in {B}elousov--{Z}habotinsky medium.
\newblock {\em Chaos, Solitons \& Fractals}, 21(5):1259--1264, 2004.

\bibitem{adamatzky2016advances}
Andrew Adamatzky, editor.
\newblock {\em Advances in Unconventional Computing: Volume 1: Theory},
  volume~22.
\newblock Springer, 2016.

\bibitem{adamatzky2011polymorphic}
Andrew Adamatzky, Ben de~Lacy~Costello, and Larry Bull.
\newblock On polymorphic logical gates in subexcitable chemical medium.
\newblock {\em International Journal of Bifurcation and Chaos},
  21(07):1977--1986, 2011.

\bibitem{adamatzky2011towards}
Andrew Adamatzky, Ben De~Lacy~Costello, Larry Bull, and Julian Holley.
\newblock Towards arithmetic circuits in sub-excitable chemical media.
\newblock {\em Israel Journal of Chemistry}, 51(1):56--66, 2011.

\bibitem{adamatzky2002collision}
Andrew Adamatzky and Benjamin de~Lacy~Costello.
\newblock Collision-free path planning in the {B}elousov-{Z}habotinsky medium
  assisted by a cellular automaton.
\newblock {\em Naturwissenschaften}, 89(10):474--478, 2002.

\bibitem{adamatzky2007binary}
Andrew Adamatzky and Benjamin de~Lacy~Costello.
\newblock Binary collisions between wave-fragments in a sub-excitable
  {B}elousov--{Z}habotinsky medium.
\newblock {\em Chaos, Solitons \& Fractals}, 34(2):307--315, 2007.

\bibitem{adamatzky2004experimental}
Andrew Adamatzky, Benjamin de~Lacy~Costello, Chris Melhuish, and Norman
  Ratcliffe.
\newblock Experimental implementation of mobile robot taxis with onboard
  {B}elousov--{Z}habotinsky chemical medium.
\newblock {\em Materials Science and Engineering: C}, 24(4):541--548, 2004.

\bibitem{adamatzky2019computing}
Andrew Adamatzky, Florian Huber, and J{\"o}rg Schnau{\ss}.
\newblock Computing on actin bundles network.
\newblock {\em Scientific reports}, 9(1):1--10, 2019.

\bibitem{adamatzky2021thoughts}
Andrew Adamatzky and Louis-Jose Lestocart, editors.
\newblock {\em Thoughts on unconventional computing}.
\newblock Luniver Press, 2021.

\bibitem{adamatzky2019actin}
Andrew Adamatzky, J{\"o}rg Schnau{\ss}, and Florian Huber.
\newblock Actin droplet machine.
\newblock {\em Royal Society Open Science}, 6(12):191135, 2019.

\bibitem{ahn2020synchrony}
Jungryul Ahn, Huu~Lam Phan, Seongkwang Cha, Kyo-in Koo, Yongseok Yoo, and
  Yong~Sook Goo.
\newblock Synchrony of spontaneous burst firing between retinal ganglion cells
  across species.
\newblock {\em Experimental Neurobiology}, 29(4):285, 2020.

\bibitem{aldridge1991temporal}
J~Wayne Aldridge and Sid Gilman.
\newblock The temporal structure of spike trains in the primate basal ganglia:
  afferent regulation of bursting demonstrated with precentral cerebral
  cortical ablation.
\newblock {\em Brain research}, 543(1):123--138, 1991.

\bibitem{antimisiaris2021overcoming}
SG~Antimisiaris, A~Marazioti, M~Kannavou, E~Natsaridis, F~Gkartziou, G~Kogkos,
  and S~Mourtas.
\newblock Overcoming barriers by local drug delivery with liposomes.
\newblock {\em Advanced Drug Delivery Reviews}, 2021.

\bibitem{beeler1977reconstruction}
Go~W Beeler and H~Reuter.
\newblock Reconstruction of the action potential of ventricular myocardial
  fibres.
\newblock {\em The Journal of physiology}, 268(1):177--210, 1977.

\bibitem{belousov1959periodic}
Boris~P {B}elousov.
\newblock A periodic reaction and its mechanism.
\newblock {\em Compilation of Abstracts on Radiation Medicine}, 147(145):1,
  1959.

\bibitem{bi1994evidence}
YU~Bi, Aristotel Pappelis, C~Steven Sikes, and Sidney~W Fox.
\newblock Evidence that the protocell was also a protoneuron.
\newblock 1994.

\bibitem{cans2003artificial}
Ann-Sofie Cans, Nathan Wittenberg, Roger Karlsson, Leslie Sombers, Mattias
  Karlsson, Owe Orwar, and Andrew Ewing.
\newblock Artificial cells: unique insights into exocytosis using liposomes and
  lipid nanotubes.
\newblock {\em Proceedings of the National Academy of Sciences},
  100(2):400--404, 2003.

\bibitem{cenci2021eco}
Marcelo~Pilotto Cenci, Tatiana Scarazzato, Daniel~Dotto Munchen, Paula~Cristina
  Dartora, Hugo~Marcelo Veit, Andrea~Moura Bernardes, and Pablo~R Dias.
\newblock Eco-friendly electronics—a comprehensive review.
\newblock {\em Advanced Materials Technologies}, page 2001263, 2021.

\bibitem{chang2017circuits}
Joseph~S Chang, Antonio~F Facchetti, and Robert Reuss.
\newblock A circuits and systems perspective of organic/printed electronics:
  review, challenges, and contemporary and emerging design approaches.
\newblock {\em IEEE Journal on emerging and selected topics in circuits and
  systems}, 7(1):7--26, 2017.

\bibitem{cocatre1992identification}
JH~Cocatre-Zilgien and F~Delcomyn.
\newblock Identification of bursts in spike trains.
\newblock {\em Journal of neuroscience methods}, 41(1):19--30, 1992.

\bibitem{dale2017reservoir}
Matthew Dale, Julian~F Miller, and Susan Stepney.
\newblock Reservoir computing as a model for in-materio computing.
\newblock In {\em Advances in Unconventional Computing}, pages 533--571.
  Springer, 2017.

\bibitem{dale2019substrate}
Matthew Dale, Julian~F Miller, Susan Stepney, and Martin~A Trefzer.
\newblock A substrate-independent framework to characterize reservoir
  computers.
\newblock {\em Proceedings of the Royal Society A}, 475(2226):20180723, 2019.

\bibitem{dehshibi2021electrical}
Mohammad~Mahdi Dehshibi and Andrew Adamatzky.
\newblock Electrical activity of fungi: Spikes detection and complexity
  analysis.
\newblock {\em Biosystems}, 203:104373, 2021.

\bibitem{desai1999plasticity}
Niraj~S Desai, Lana~C Rutherford, and Gina~G Turrigiano.
\newblock Plasticity in the intrinsic excitability of cortical pyramidal
  neurons.
\newblock {\em Nature neuroscience}, 2(6):515--520, 1999.

\bibitem{dorval2008probability}
Alan~D Dorval.
\newblock Probability distributions of the logarithm of inter-spike intervals
  yield accurate entropy estimates from small datasets.
\newblock {\em Journal of neuroscience methods}, 173(1):129--139, 2008.

\bibitem{egorov2021robotics}
Egor Egorov, Calvin Pieters, Hila Korach-Rechtman, Jeny Shklover, and Avi
  Schroeder.
\newblock Robotics, microfluidics, nanotechnology and ai in the synthesis and
  evaluation of liposomes and polymeric drug delivery systems.
\newblock {\em Drug Delivery and Translational Research}, 11(2):345--352, 2021.

\bibitem{fahlman2019interfaces}
Mats Fahlman, Simone Fabiano, Viktor Gueskine, Daniel Simon, Magnus Berggren,
  and Xavier Crispin.
\newblock Interfaces in organic electronics.
\newblock {\em Nature Reviews Materials}, 4(10):627--650, 2019.

\bibitem{feron2018organic}
Krishna Feron, Rebecca Lim, Connor Sherwood, Angela Keynes, Alan Brichta, and
  Paul~C Dastoor.
\newblock Organic bioelectronics: materials and biocompatibility.
\newblock {\em International journal of molecular sciences}, 19(8):2382, 2018.

\bibitem{fitzhugh1961impulses}
Richard FitzHugh.
\newblock Impulses and physiological states in theoretical models of nerve
  membrane.
\newblock {\em Biophysical journal}, 1(6):445--466, 1961.

\bibitem{follmann1982deoxyribonucleotide}
Hartmut Follmann.
\newblock Deoxyribonucleotide synthesis and the emergence of dna in molecular
  evolution.
\newblock {\em Naturwissenschaften}, 69(2):75--81, 1982.

\bibitem{fox1992thermal}
Sidney~W Fox.
\newblock Thermal proteins in the first life and in the ``mind-body” problem.
\newblock In {\em Evolution of Information Processing Systems}, pages 203--228.
  Springer, 1992.

\bibitem{fox1995experimental}
Sidney~W Fox, Peter~R Bahn, Klaus Dose, Kaoru Harada, Laura Hsu, Yoshio Ishima,
  John Jungck, Jean Kendrick, Gottfried Krampitz, James~C Lacey, et~al.
\newblock Experimental retracement of the origins of a protocell.
\newblock {\em Journal of biological physics}, 20(1-4):17--36, 1995.

\bibitem{friederich2019toward}
Pascal Friederich, Artem Fediai, Simon Kaiser, Manuel Konrad, Nicole Jung, and
  Wolfgang Wenzel.
\newblock Toward design of novel materials for organic electronics.
\newblock {\em Advanced Materials}, 31(26):1808256, 2019.

\bibitem{gentili2012belousov}
Pier~Luigi Gentili, Viktor Horvath, Vladimir~K Vanag, and Irving~R Epstein.
\newblock Belousov-{Z}habotinsky ``chemical neuron'' as a binary and fuzzy
  logic processor.
\newblock {\em IJUC}, 8(2):177--192, 2012.

\bibitem{gorecki2014information}
J~Gorecki, JN~Gorecka, and Andrew Adamatzky.
\newblock Information coding with frequency of oscillations in
  {B}elousov-{Z}habotinsky encapsulated disks.
\newblock {\em Physical Review E}, 89(4):042910, 2014.

\bibitem{gorecki2003chemical}
J~Gorecki, K~Yoshikawa, and Y~Igarashi.
\newblock On chemical reactors that can count.
\newblock {\em The Journal of Physical Chemistry A}, 107(10):1664--1669, 2003.

\bibitem{gorecki2006information}
Jerzy Gorecki and Joanna~Natalia Gorecka.
\newblock Information processing with chemical excitations--from instant
  machines to an artificial chemical brain.
\newblock {\em International Journal of Unconventional Computing}, 2(4), 2006.

\bibitem{gorecki2009information}
Jerzy Gorecki, Joanna~Natalia Gorecka, and Yasuhiro Igarashi.
\newblock Information processing with structured excitable medium.
\newblock {\em Natural Computing}, 8(3):473--492, 2009.

\bibitem{gruenert2015understanding}
Gerd Gruenert, Konrad Gizynski, Gabi Escuela, Bashar Ibrahim, Jerzy Gorecki,
  and Peter Dittrich.
\newblock Understanding networks of computing chemical droplet neurons based on
  information flow.
\newblock {\em International journal of neural systems}, 25(07):1450032, 2015.

\bibitem{han2020advanced}
Won~Bae Han, Joong~Hoon Lee, Jeong-Woong Shin, and Suk-Won Hwang.
\newblock Advanced materials and systems for biodegradable, transient
  electronics.
\newblock {\em Advanced Materials}, 32(51):2002211, 2020.

\bibitem{harada1958thermal}
Kaoru Harada and Sidney~W Fox.
\newblock The thermal condensation of glutamic acid and glycine to linear
  peptides1.
\newblock {\em Journal of the American Chemical Society}, 80(11):2694--2697,
  1958.

\bibitem{hsu1971conjugation}
Laura~Ling Hsu, Steven Brooke, and Sidney~W Fox.
\newblock Conjugation of proteinoid microspheres: a model of primordial
  communication.
\newblock {\em Biosystems}, 4(1):12--25, 1971.

\bibitem{Nomura2020}
Shin ichiro Nomura, Gen Hayase, Taro Toyota, Richard Mayne, and Andrew
  Adamatzky.
\newblock {On Multicellular Lipid Compartments and Their Electrical Activity}.
\newblock {\em 10.26434/chemrxiv.12129114.v1}, 4 2020.

\bibitem{DBLP:journals/ijuc/IgarashiG11}
Yasuhiro Igarashi and Jerzy Gorecki.
\newblock Chemical diodes built with controlled excitable media.
\newblock {\em {IJUC}}, 7(3):141--158, 2011.

\bibitem{ishima1981electrical}
Yoshio Ishima, Aleksander~T Przybylski, and Sidney~W Fox.
\newblock Electrical membrane phenomena in spherules from proteinoid and
  lecithin.
\newblock {\em BioSystems}, 13(4):243--251, 1981.

\bibitem{ji2019recent}
Deyang Ji, Tao Li, Wenping Hu, and Harald Fuchs.
\newblock Recent progress in aromatic polyimide dielectrics for organic
  electronic devices and circuits.
\newblock {\em Advanced Materials}, 31(15):1806070, 2019.

\bibitem{jun2020role}
Yuju Jun, Hyunyoung Oh, Rajshekhar Karpoormath, Amitabh Jha, and Rajkumar
  Patel.
\newblock Role of microsphere as drug carrier for osteogenic differentiation.
\newblock {\em International Journal of Polymeric Materials and Polymeric
  Biomaterials}, pages 1--10, 2020.

\bibitem{kaminaga2006reaction}
Akiko Kaminaga, Vladimir~K Vanag, and Irving~R Epstein.
\newblock A reaction--diffusion memory device.
\newblock {\em Angewandte Chemie International Edition}, 45(19):3087--3089,
  2006.

\bibitem{kamiya2017giant}
Koki Kamiya and Shoji Takeuchi.
\newblock Giant liposome formation toward the synthesis of well-defined
  artificial cells.
\newblock {\em Journal of Materials Chemistry B}, 5(30):5911--5923, 2017.

\bibitem{kimizuka1964ion}
H~Kimizuka and K~Koketsu.
\newblock Ion transport through cell membrane.
\newblock {\em Journal of Theoretical Biology}, 6(2):290--305, 1964.

\bibitem{kokufuta1983factors}
Etsuo Kokufuta, Hidetoshi Sakai, and Kaoru Harada.
\newblock Factors controlling the size of proteinoid microspheres.
\newblock {\em BioSystems}, 16(3-4):175--181, 1983.

\bibitem{kolitz2014engineering}
Michal Kolitz-Domb, Igor Grinberg, Enav Corem-Salkmon, and Shlomo Margel.
\newblock Engineering of near infrared fluorescent proteinoid-poly (l-lactic
  acid) particles for in vivo colon cancer detection.
\newblock {\em Journal of nanobiotechnology}, 12(1):30, 2014.

\bibitem{kolitz2015engineering}
Michal Kolitz-Domb and Shlomo Margel.
\newblock Engineering of novel proteinoids and plla-proteinoid polymers of
  narrow size distribution and uniform nano/micro-hollow particles for
  biomedical applications.
\newblock In {\em Advances in Bioengineering}. IntechOpen, 2015.

\bibitem{kolitz2018recent}
Michal Kolitz-Domb and Shlomo Margel.
\newblock Recent advances of novel proteinoids and proteinoid nanoparticles and
  their applications in biomedicine and industrial uses.
\newblock {\em Israel Journal of Chemistry}, 58(12):1277--1285, 2018.

\bibitem{konkoli2018reservoir}
Zoran Konkoli, Stefano Nichele, Matthew Dale, and Susan Stepney.
\newblock Reservoir computing with computational matter.
\newblock In {\em Computational Matter}, pages 269--293. Springer, 2018.

\bibitem{kuhnert1986new}
L~Kuhnert.
\newblock A new optical photochemical memory device in a light-sensitive
  chemical active medium.
\newblock 1986.

\bibitem{kuhnert1989image}
Lothar Kuhnert, KI~Agladze, and VI~Krinsky.
\newblock Image processing using light-sensitive chemical waves.
\newblock 1989.

\bibitem{kumar1998preparation}
AB~Madhan Kumar and K~Panduranga Rao.
\newblock Preparation and characterization of ph-sensitive proteinoid
  microspheres for the oral delivery of methotrexate.
\newblock {\em Biomaterials}, 19(7-9):725--732, 1998.

\bibitem{lee2017toward}
Eun~Kwang Lee, Moo~Yeol Lee, Cheol~Hee Park, Hae~Rang Lee, and Joon~Hak Oh.
\newblock Toward environmentally robust organic electronics: approaches and
  applications.
\newblock {\em Advanced Materials}, 29(44):1703638, 2017.

\bibitem{li2020biodegradable}
Wenhui Li, Qian Liu, Yuniu Zhang, Chang'an Li, Zhenfei He, Wallace~CH Choy,
  Paul~J Low, Prashant Sonar, and Aung Ko~Ko Kyaw.
\newblock Biodegradable materials and green processing for green electronics.
\newblock {\em Advanced Materials}, 32(33):2001591, 2020.

\bibitem{litzinger1992phosphatodylethanolamine}
David~C Litzinger and Leaf Huang.
\newblock Phosphatodylethanolamine liposomes: drug delivery, gene transfer and
  immunodiagnostic applications.
\newblock {\em Biochimica et Biophysica Acta (BBA)-Reviews on Biomembranes},
  1113(2):201--227, 1992.

\bibitem{lobov2020competitive}
Sergey~A Lobov, Andrey~V Chernyshov, Nadia~P Krilova, Maxim~O Shamshin, and
  Victor~B Kazantsev.
\newblock Competitive learning in a spiking neural network: towards an
  intelligent pattern classifier.
\newblock {\em Sensors}, 20(2):500, 2020.

\bibitem{lukovsevivcius2009reservoir}
Mantas Luko{\v{s}}evi{\v{c}}ius and Herbert Jaeger.
\newblock Reservoir computing approaches to recurrent neural network training.
\newblock {\em Computer Science Review}, 3(3):127--149, 2009.

\bibitem{mao2019bio}
Jing-Yu Mao, Li~Zhou, Yi~Ren, Jia-Qin Yang, Chih-Li Chang, Heng-Chuan Lin,
  Ho-Hsiu Chou, Shi-Rui Zhang, Ye~Zhou, and Su-Ting Han.
\newblock A bio-inspired electronic synapse using solution processable organic
  small molecule.
\newblock {\em Journal of Materials Chemistry C}, 7(6):1491--1501, 2019.

\bibitem{matsui2019flexible}
Hiroyuki Matsui, Yasunori Takeda, and Shizuo Tokito.
\newblock Flexible and printed organic transistors: From materials to
  integrated circuits.
\newblock {\em Organic Electronics}, 75:105432, 2019.

\bibitem{matsuno1984electrical}
Koichiro Matsuno.
\newblock Electrical excitability of proteinoid microspheres composed of basic
  and acidic proteinoids.
\newblock {\em BioSystems}, 17(1):11--14, 1984.

\bibitem{matsuno2012molecular}
Koichiro Matsuno.
\newblock {\em Molecular Evolution and Protobiology}.
\newblock Springer Science \& Business Media, 2012.

\bibitem{meierhenrich2010origin}
Uwe~J Meierhenrich, Jean-Jacques Filippi, Cornelia Meinert, Pierre Vierling,
  and Jason~P Dworkin.
\newblock On the origin of primitive cells: from nutrient intake to elongation
  of encapsulated nucleotides.
\newblock {\em Angewandte Chemie International Edition}, 49(22):3738--3750,
  2010.

\bibitem{miller2002evolution}
Julian~F Miller and Keith Downing.
\newblock Evolution in materio: Looking beyond the silicon box.
\newblock In {\em Proceedings 2002 NASA/DoD Conference on Evolvable Hardware},
  pages 167--176. IEEE, 2002.

\bibitem{miller2014evolution}
Julian~F Miller, Simon~L Harding, and Gunnar Tufte.
\newblock Evolution-in-materio: evolving computation in materials.
\newblock {\em Evolutionary Intelligence}, 7(1):49--67, 2014.

\bibitem{miller2018materio}
Julian~F Miller, Simon~J Hickinbotham, and Martyn Amos.
\newblock In materio computation using carbon nanotubes.
\newblock In {\em Computational Matter}, pages 33--43. Springer, 2018.

\bibitem{miller2019alchemy}
Julian~Francis Miller.
\newblock The alchemy of computation: designing with the unknown.
\newblock {\em Natural Computing}, 18(3):515--526, 2019.

\bibitem{nagumo1962active}
Jinichi Nagumo, Suguru Arimoto, and Shuji Yoshizawa.
\newblock An active pulse transmission line simulating nerve axon.
\newblock {\em Proceedings of the IRE}, 50(10):2061--2070, 1962.

\bibitem{nakata1998self}
Statoshi Nakata, Takahiro Miyata, Nozomi Ojima, and Kenichi Yoshikawa.
\newblock Self-synchronization in coupled salt-water oscillators.
\newblock {\em Physica D: Nonlinear Phenomena}, 115(3-4):313--320, 1998.

\bibitem{nielsen2009biomimetic}
Claus~H{\'e}lix Nielsen.
\newblock Biomimetic membranes for sensor and separation applications.
\newblock {\em Analytical and bioanalytical chemistry}, 395(3):697--718, 2009.

\bibitem{perkel1967neuronal}
Donald~H Perkel, George~L Gerstein, and George~P Moore.
\newblock Neuronal spike trains and stochastic point processes: I. the single
  spike train.
\newblock {\em Biophysical journal}, 7(4):391--418, 1967.

\bibitem{pertsov1993spiral}
Arkady~M Pertsov, Jorge~M Davidenko, Remy Salomonsz, William~T Baxter, and Jose
  Jalife.
\newblock Spiral waves of excitation underlie reentrant activity in isolated
  cardiac muscle.
\newblock {\em Circulation research}, 72(3):631--650, 1993.

\bibitem{pohorille2002artificial}
Andrew Pohorille and David Deamer.
\newblock Artificial cells: prospects for biotechnology.
\newblock {\em Trends in biotechnology}, 20(3):123--128, 2002.

\bibitem{privman2018enzyme}
Marina Privman, Nataliia Guz, and Evgeny Katz.
\newblock Enzyme-logic digital biosensors for biomedical applications.
\newblock {\em International Journal of Unconventional Computing}, 13(6), 2018.

\bibitem{przybylski1984physical}
Aleksander~T Przybylski.
\newblock Physical background of excitability: synthetic membranes and
  excitable cells.
\newblock In {\em Molecular Evolution and Protobiology}, pages 253--266.
  Springer, 1984.

\bibitem{przybylski1985excitable}
Aleksander~T Przybylski.
\newblock Excitable cell made of thermal proteinoids.
\newblock {\em BioSystems}, 17(4):281--288, 1985.

\bibitem{przybylski1982membrane}
Aleksander~T Przybylski, Wilford~P Stratten, Robert~M Syren, and Sidney~W Fox.
\newblock Membrane, action, and oscillatory potentials in simulated protocells.
\newblock {\em Naturwissenschaften}, 69(12):561--563, 1982.

\bibitem{przybylski1983towards}
AT~Przybylski, RM~Syren, and SW~Fox.
\newblock Towards an organic photobattery-photovoltaic properties of some
  thermal copolyamino acids.
\newblock 1983.

\bibitem{quirk2010triggered}
Stephen Quirk.
\newblock Triggered release from peptide-proteinoid microspheres.
\newblock {\em Journal of Biomedical Materials Research Part A},
  92(3):877--886, 2010.

\bibitem{rambidi2001chemical}
NG~Rambidi and D~Yakovenchuk.
\newblock Chemical reaction-diffusion implementation of finding the shortest
  paths in a labyrinth.
\newblock {\em Physical Review E}, 63(2):026607, 2001.

\bibitem{rizzotti1998did}
M~Rizzotti, M~Crisma, F~De~Luca, P~Iobstraibizer, and P~Mazzei.
\newblock Did the first cell emerge from a microsphere?
\newblock In {\em Exobiology: Matter, Energy, and Information in the Origin and
  Evolution of Life in the Universe}, pages 199--202. Springer, 1998.

\bibitem{Rohlfing1970998}
D.L. Rohlfing.
\newblock Catalytic activities of thermally prepared polyo-Î±-amino acids:
  Effect of aging.
\newblock {\em Science}, 169(3949):998--1000, 1970.

\bibitem{sasson2020engineering}
Elisheva Sasson, Ruth Van~Oss Pinhasi, Shlomo Margel, and Liron Klipcan.
\newblock Engineering and use of proteinoid polymers and nanocapsules
  containing agrochemicals.
\newblock {\em Scientific reports}, 10(1):1--13, 2020.

\bibitem{schreiber2019prebiotic}
Andreas Schreiber, Matthias~C Huber, and Stefan~M Schiller.
\newblock Prebiotic protocell model based on dynamic protein membranes
  accommodating anabolic reactions.
\newblock {\em Langmuir}, 35(29):9593--9610, 2019.

\bibitem{sharov2016coenzyme}
Alexei~A Sharov.
\newblock Coenzyme world model of the origin of life.
\newblock {\em Biosystems}, 144:8--17, 2016.

\bibitem{sielewiesiuk2001logical}
Jakub Sielewiesiuk and Jerzy G{\'o}recki.
\newblock Logical functions of a cross junction of excitable chemical media.
\newblock {\em The Journal of Physical Chemistry A}, 105(35):8189--8195, 2001.

\bibitem{slomowitz2015interplay}
Edden Slomowitz, Boaz Styr, Irena Vertkin, Hila Milshtein-Parush, Israel
  Nelken, Michael Slutsky, and Inna Slutsky.
\newblock Interplay between population firing stability and single neuron
  dynamics in hippocampal networks.
\newblock {\em Elife}, 4:e04378, 2015.

\bibitem{stano2017minimal}
Pasquale Stano.
\newblock Minimal cellular models for origins-of-life studies and
  biotechnology.
\newblock In {\em The Biophysics of Cell Membranes}, pages 177--219. Springer,
  2017.

\bibitem{steinbock1996chemical}
Oliver Steinbock, Petteri Kettunen, and Kenneth Showalter.
\newblock Chemical wave logic gates.
\newblock {\em The Journal of Physical Chemistry}, 100(49):18970--18975, 1996.

\bibitem{steinbock1995navigating}
Oliver Steinbock, {\'A}gota T{\'o}th, and Kenneth Showalter.
\newblock Navigating complex labyrinths: optimal paths from chemical waves.
\newblock {\em Science}, pages 868--868, 1995.

\bibitem{stepney2019co}
Susan Stepney.
\newblock Co-designing the computational model and the computing substrate.
\newblock In {\em International Conference on Unconventional Computation and
  Natural Computation}, pages 5--14. Springer, 2019.

\bibitem{stevens2012time}
William~M Stevens, Andrew Adamatzky, Ishrat Jahan, and Ben de~Lacy~Costello.
\newblock Time-dependent wave selection for information processing in excitable
  media.
\newblock {\em Physical Review E}, 85(6):066129, 2012.

\bibitem{stovold2012simulating}
James Stovold and Simon O'Keefe.
\newblock Simulating neurons in reaction-diffusion chemistry.
\newblock In {\em International Conference on Information Processing in Cells
  and Tissues}, pages 143--149. Springer, 2012.

\bibitem{stovold2016reaction}
James Stovold and Simon O'Keefe.
\newblock Reaction--diffusion chemistry implementation of associative memory
  neural network.
\newblock {\em International Journal of Parallel, Emergent and Distributed
  Systems}, pages 1--21, 2016.

\bibitem{stovold2017associative}
James Stovold and Simon O'Keefe.
\newblock Associative memory in reaction-diffusion chemistry.
\newblock In {\em Advances in Unconventional Computing}, pages 141--166.
  Springer, 2017.

\bibitem{sun2011burst}
Xiaojuan Sun, Jinzhi Lei, Matja{\v{z}} Perc, J{\"u}rgen Kurths, and Guanrong
  Chen.
\newblock Burst synchronization transitions in a neuronal network of
  subnetworks.
\newblock {\em Chaos: An Interdisciplinary Journal of Nonlinear Science},
  21(1):016110, 2011.

\bibitem{takigawa2011dendritic}
Hisako Takigawa-Imamura and Ikuko~N Motoike.
\newblock Dendritic gates for signal integration with excitability-dependent
  responsiveness.
\newblock {\em Neural Networks}, 24(10):1143--1152, 2011.

\bibitem{tallawi2011proteinoid}
Marwa Tallawi.
\newblock Proteinoid/hydroxyapatite hybrid microsphere composites.
\newblock {\em Journal of Biomedical Materials Research Part B: Applied
  Biomaterials}, 96(2):261--266, 2011.

\bibitem{tamagawa2015membrane}
Hirohisa Tamagawa.
\newblock Membrane potential generation without ion transport.
\newblock {\em Ionics}, 21(6):1631--1648, 2015.

\bibitem{toth2010simple}
Rita Toth, Christopher Stone, Ben de~Lacy~Costello, Andrew Adamatzky, and Larry
  Bull.
\newblock Simple collision-based chemical logic gates with adaptive computing.
\newblock {\em Theoretical and Technological Advancements in Nanotechnology and
  Molecular Computation: Interdisciplinary Gains: Interdisciplinary Gains},
  page 162, 2010.

\bibitem{tsakalos2021protein}
Karolos-Alexandros Tsakalos, Georgios~Ch Sirakoulis, Andy Adamatzky, and Jim
  Smith.
\newblock Protein structured reservoir computing for spike-based pattern
  recognition.
\newblock {\em IEEE Transactions on Parallel and Distributed Systems}, 2021.

\bibitem{van2018organic}
Yoeri van De~Burgt, Armantas Melianas, Scott~Tom Keene, George Malliaras, and
  Alberto Salleo.
\newblock Organic electronics for neuromorphic computing.
\newblock {\em Nature Electronics}, 1(7):386--397, 2018.

\bibitem{van2017origin}
Jordi Van~Gestel and Corina~E Tarnita.
\newblock On the origin of biological construction, with a focus on
  multicellularity.
\newblock {\em Proceedings of the National Academy of Sciences},
  114(42):11018--11026, 2017.

\bibitem{vaughan1987thermal}
Graham Vaughan, Alexander~T Przybylski, and Sidney~W Fox.
\newblock Thermal proteinoids as excitability-inducing materials.
\newblock {\em BioSystems}, 20(3):219--223, 1987.

\bibitem{DBLP:journals/ijuc/Vazquez-OteroFDD14}
Alejandro Vazquez{-}Otero, Jan Faigl, Natividad Duro, and Raquel Dormido.
\newblock Reaction-diffusion based computational model for autonomous mobile
  robot exploration of unknown environments.
\newblock {\em {IJUC}}, 10(4):295--316, 2014.

\bibitem{verstraeten2007experimental}
David Verstraeten, Benjamin Schrauwen, Michiel d’Haene, and Dirk Stroobandt.
\newblock An experimental unification of reservoir computing methods.
\newblock {\em Neural networks}, 20(3):391--403, 2007.

\bibitem{wills2020metrics}
Peter Wills and Fran{\c{c}}ois~G Meyer.
\newblock Metrics for graph comparison: A practitioner’s guide.
\newblock {\em Plos one}, 15(2):e0228728, 2020.

\bibitem{wu2021biodegradable}
Wei Wu.
\newblock Biodegradable polymer nanocomposites for electronics.
\newblock {\em Polymer Nanocomposite Materials: Applications in Integrated
  Electronic Devices}, pages 53--75, 2021.

\bibitem{yokoi2004excitable}
Hiroshi Yokoi, Andy Adamatzky, Ben de~Lacy~Costello, and Chris Melhuish.
\newblock Excitable chemical medium controller for a robotic hand: Closed-loop
  experiments.
\newblock {\em International Journal of Bifurcation and Chaos},
  14(09):3347--3354, 2004.

\bibitem{zavalov2012enzyme}
Oleksandr Zavalov, Vera Bocharova, Vladimir Privman, and Evgeny Katz.
\newblock Enzyme-based logic: Or gate with double-sigmoid filter response.
\newblock {\em The Journal of Physical Chemistry B}, 116(32):9683--9689, 2012.

\bibitem{zhabotinsky1964periodic}
AM~{Z}habotinsky.
\newblock Periodic processes of malonic acid oxidation in a liquid phase.
\newblock {\em Biofizika}, 9(306-311):11, 1964.

\bibitem{zhang2021fabrication}
Xunan Zhang, Xiaotong Shao, Zhenzhen Cai, Xinyu Yan, and Wei Zong.
\newblock The fabrication of phospholipid vesicle-based artificial cells and
  their functions.
\newblock {\em New Journal of Chemistry}, 45(7):3364--3376, 2021.

\bibitem{zhang2008artificial}
Ying Zhang, Warren~C Ruder, and Philip~R LeDuc.
\newblock Artificial cells: building bioinspired systems using small-scale
  biology.
\newblock {\em Trends in biotechnology}, 26(1):14--20, 2008.

\bibitem{zheng2008spatiotemporal}
Yan~Hong Zheng and Qi~Shao Lu.
\newblock Spatiotemporal patterns and chaotic burst synchronization in a
  small-world neuronal network.
\newblock {\em Physica A: Statistical Mechanics and its Applications},
  387(14):3719--3728, 2008.

\end{thebibliography}


\section*{References}

\end{document}